\documentstyle[onecolumn,epsf]{mn}

\newcommand{\ba}{\begin{eqnarray}}
\newcommand{\ea}{\end{eqnarray}}
\newcommand{\be}{\begin{equation}}
\newcommand{\ee}{\end{equation}}
\newcommand{\bv}{{\bf v}}
\newcommand{\bu}{{\bf u}}
\newcommand{\bA}{{\bf A}}
\newcommand{\tomega}{{\tilde\omega}}
\newcommand{\rmRe}{{\rm Re}}
\newcommand{\rmIm}{{\rm Im}}
\def\go{\mathrel{\raise.3ex\hbox{$>$}\mkern-14mu
             \lower0.6ex\hbox{$\sim$}}}
\def\lo{\mathrel{\raise.3ex\hbox{$<$}\mkern-14mu
             \lower0.6ex\hbox{$\sim$}}}

\title[Wave Excitation in Three-Dimensional Disks by External Potential]
{Wave Excitation in Three-Dimensional Disks by External Potential}

\author[Hang Zhang and Dong Lai]{Hang Zhang$^{1,2}$\thanks{Email:
zhanghang@njnu.edu.cn; dong@astro.cornell.edu} and Dong
Lai$^{2,3}$\footnotemark[1] \\ $^1$Department of Physics, Nanjing Normal
University, Nanjing, JS 210097, P.R.~China\\ $^2$Department of Astronomy,
Cornell University, Ithaca, NY 14853, USA \\
$^3$National Astronomical Observatories, Chinese Academy of
Sciences, Beijing, 100012, China\\}

\begin{document}

\date{Accepted ????. Received ?????; in original form ?????}

\pagerange{\pageref{firstpage}--\pageref{lastpage}} \pubyear{2005}

\maketitle

\label{firstpage}

\begin{abstract}
We study the excitation of density and bending waves and the associated 
angular momentum transfer in gaseous disks with finite thickness by a 
rotating external potential. The disk is assumed to be isothermal in the 
vertical direction and has no self-gravity. The disk perturbations are 
decomposed into different modes, each characterized by the azimuthal 
index $m$ and the vertical index $n$, which specifies the nodal number 
of the density perturbation along the disk normal direction.
The $n=0$ modes correspond to the two-dimensional density waves
previously studied by Goldreich \& Tremaine and others.
In a three-dimensional disk, waves can be excited at both
Lindblad resonances (for modes with $n=0,1,2,\cdots$) and
vertical resonances (for the $n\ge 1$ modes only).
The torque on the disk is positive for waves excited at outer
Linblad/vertical resonances and negative at inner Lindblad/vertical
resonances. While the $n=0$ modes are evanescent around corotation, 
the $n\ge 1$ modes can propagate into the corotation region where 
they are damped and deposit their angular momenta.
We have derived analytical expressions for the amplitudes of different wave
modes excited at Lindblad and/or vertical resonances and the resulting 
torques on the disk. It is found that for $n\ge 1$, angular momentum
transfer through vertical resonances is much more efficient than
Lindblad resonances. This implies that in some situations
(e.g., a circumstellar disk perturbed by a planet in an inclined orbit),
vertical resonances may be an important channel of angular momentum
transfer between the disk and the external potential.
We have also derived new formulae for the angular momentum 
deposition at corotation and studied wave excitations at disk boundaries.
\end{abstract}


\section{Introduction}

Gravitational interaction between a gaseous disk and an an external body plays 
an important role in many astrophysical systems, including protoplanetary
disks, binary stars and spiral galaxies.
The external potential generates
disturbances in the disk, shaping the structure and evolution of the
disk, and these, in turn, influence the dynamics of the external
object itself.  In a classic paper, Goldreich \& Tremaine (1979;
hereafter GT) studied density wave excitation by an external potential
in a two-dimensional disk and gave the formulation of angular
momentum transport rate or torque at Lindblad and corotation
resonances for disks with or without self-gravity (see 
also Goldreich \& Tremaine 1978; Lin \& Papaloizou 1979). Since then,
numerous extensions and applications of their theory have appeared in
the literature. For example, Shu, Yuan \& Lissauer (1985)
studied the effect of nonlinearities of the waves on the resonant torque.  
Meyer-Vernet \& Sicardy (1987) examined the transfer
of angular momentum in a disk subjected to perturbations at Lindblad resonance
under various physical conditions.  
Artymowicz (1993) derived a generalized torque formula which
is useful for large azimuthal numbers.  The saturation of the
corotation resonance were investigated in detail by many authors (e.g., 
Balmforth \& Korycansky 2001; Ogilvie \& Lubow
2003). Korycansky \& Pollack (1993) performed numerical calculations
of the torques. Applications of the GT theory were mostly focused on
disk-satellite interactions, including the eccentricity and inclination
evolution of the satellite's orbit (e.g., Goldreich \& Tremaine 1980; Ward
1988; Goldreich \& Sari 2003)
and protoplanet migration in the Solar Nebula (e.g., Ward 1986,~1997; 
Masset \& Papaloizou 2003).

A number of studies have been devoted to the three-dimensional (3D)
responses of a disk to an external potential.  Lubow (1981) analyzed
wave generation by tidal force at the vertical resonance in an
isothermal accretion disk and investigated the possibility that the
resonantly driven waves can maintain a self-sustained accretion.
Ogilvie (2002) generalized Lubow's analysis to non-isothermal disks
and also considered nonlinear effects.  Lubow \& Pringle (1993)
studied the propagation property of 3D axisymmetric waves in
disks, and Bate et al.~(2002) examined the excitation, propagation and 
dissipation of axisymmetric waves.
Wave excitation at Lindblad resonances in thermally stratified
disks was investigated by Lubow \& Ogilvie (1998) using shearing sheet
model. Takeuchi \& Miyama (1998) studied wave generation at vertical
resonances in isothermal disks by external gravity.  Tanaka, Takeuchi
\& Ward (2002) investigated the corotation and Lindblad torque brought
forth from the 3D interaction between a planet and an isothermal
gaseous disk.  The excitation of bending waves was studied by
Papaloizou \& Lin (1995) and Terquem (1998) in disks without
resonances.  Artymowicz (1994), Ward \& Hahn (2003) and Tanaka \& Ward
(2004) investigated many aspects of resonance-driven bending waves.

Despite all these fruitful researches, a unified description and
analysis of wave excitation in 3D disks by external potentials are
still desirable. This is the purpose of our paper.
As a first step, we only consider linear theory.  Our treatment allows
for Lindblad, corotation and vertical resonances to be described within
the same theoretical framework, and in the meantime, density waves and
bending waves to be handled in an unified manner. By taking advantage
of Fourier-Hermite expansion, different modes of perturbation for
locally isothermal disks are well organized and a second-order
differential equation for individual mode is attained (see Tanaka,
Takeuchi \& Ward 2002). In order to treat it mathematically, the
derived equation which appears monstrous is pruned under different
situations for those modes with the highest order of magnitude. Then,
following the standard technique used by Goldreich \& Tremaine (1979),
the simplified equations are solved and analytic expressions for the
waves excited at various locations and associated angular momentum
transfer rates are calculated.

Our paper is organized as follows. The derivation of the basic
equations describing the response of a disk to an external
potential are briefly presented in \S 2. The assumptions 
underlying our formulation are also specified there. 
In \S3 we examine the dispersion relation for free waves and 
various resonances which may exist in the disk. Wave modes are 
organized according to the azimuthal index $m$ and the vertical index $n$, 
with $n=0$ corresponding to the modes in a 2D disk. In \S 4 we study
Lindblad resonances. \S 4.1 deals with the $n\neq1$ cases, where
the solutions pertaining to the waves excited at the resonances are found
and the angular momentum transports by such waves are
calculated. The $n=1$ case is treated separately in \S 4.2,
because in this case the Lindblad resonance coincides with the 
vertical resonance for a Keplerian disk. In \S 5 we study
wave excitation and angular momentum transport at vertical resonances
for general $n$. Corotation resonances are examined in \S 6, where
it is necessary to treat the $n\geq1$ and $n=0$ cases separately.
In \S 7 we study wave excitation at disk boundaries.
We discuss the effects of various assumptions adopted in our 
treatment in \S 8 and summarize our main result in \S 9. 
Appendix A contains general solutions of disk perturbations away from
resonances, and Appendix B gives a specific example of angular momentum flow
in the disk: wave excitation at a Lindblad or vertical resonance, followed
by wave propagation, and finally wave damping at corotation.

\section{Basic Equations}

We consider a geometrically thin gas disk and adopt cylindrical
coordinates $(r,\theta,z)$.  The unperturbed disk has velocity
$\bv_0=(0,r\Omega,0)$, where the angular velocity $\Omega=\Omega(r)$
is taken to be a function of $r$ alone. The disk is assumed to be
isothermal in the vertical direction 
and non-self-gravitating. 
The equation of hydrostatic
equilibrium (for $z\ll r$) reads 
\be 
{dp_0\over dz}=-\rho_0\Omega_\perp^2 z.  
\ee 
Thus the vertical density profile is
given by 
\be 
\rho_0(r,z)={\sigma\over\sqrt{2\pi}h}\exp (-Z^2/2),\quad
{\rm with}~~ Z=z/h 
\label{eq:rho0}\ee 
where $h=h(r)=c/\Omega_\perp$ is the disk scale
height, $c=c(r)=\sqrt{p_0/\rho_0}$ is the isothermal sound speed,
$\sigma=\sigma(r)=\int dz\,\rho_0$ is the surface density, and
$\Omega_\perp$ is the vertical oscillation frequency of the disk.

We now consider perturbation of the disk driven by an external potential
$\phi$. The linear perturbation equations read
\ba
&&{\partial \bu\over\partial t}+(\bv_0\cdot\nabla)\bu+(\bu\cdot\nabla)\bv_0
=-{1\over\rho_0}\nabla\delta P+{\delta\rho\over\rho_0^2}\nabla p_0
-\nabla\phi,\label{eq:u}\\
&& {\partial\rho\over\partial
t}+\nabla\cdot(\rho_0\bu+\bv_0\delta\rho)=0, 
\label{eq:rho}\ea 
where $\delta\rho,~\delta P$ and $\bu=\delta\bv$ are the (Eulerian)
perturbations of density, pressure and velocity, respectively. 
Without loss of generality, each perturbation variable $X$ and the external
potential $\phi$ are assumed to have the form of a normal mode in
$\theta$ and $t$
\be
X(r,\theta,z,t)=X(r,z)\exp(im\theta-i\omega t), 
\ee
where $m$ is a nonnegative integer and consequently, $\omega$ is allowed 
to be either positive or negative, corresponding to the
prograde or retrograde wave, respectively. Note that only the real part
of the perturbation has physical meaning and we will not write 
out explicitly the dependence on $m$ and $\omega$ for 
the amplitude $X(r,z)$ and other related 
quantities in this paper. The equation for adiabatic perturbations is
$dP/dt=c_s^2d\rho/dt$, where $c_s$ is the adiabatic sound speed.
This yields
\be
-i\tomega(\delta P-c_s^2\delta\rho)=c_s^2\rho_0 \bu\cdot\bA,
\label{eq:energy}\ee
where
\be
\tomega=\omega-m\Omega
\ee
is the ``Doppler-shifted'' frequency, and
\be
\bA={\nabla\rho_0\over\rho_0}-{\nabla p_0\over c_s^2\rho_0}
\label{schwarz}\ee
is the Schwarzschild discriminant vector.
In general, $|A_r|\sim 1/r$ may be neglected compared to
$|A_z|\sim 1/h$ for thin disks ($h\ll r$). In the following,
we will also assume $A_z=0$, i.e., the disk is neutrally stratified
in the vertical direction. This amounts to assuming $c_s=c$
(i.e., the perturbations are assumed isothermal).
Equation (\ref{eq:energy}) then becomes $\delta P=c^2\delta\rho$.
Introducing the enthalpy perturbation
\be
\eta=\delta p/\rho_0,
\ee
Equations (\ref{eq:u}) and (\ref{eq:rho}) reduce to
\footnote{On the right-hand-side of eq.~(\ref{eq:fluid1}), 
we have dropped the term $2(d\ln c/dr)\eta$. In effect, we assume
$c$ is constant in radius. Relaxing this assumption does not change the
results of the paper. See \S 8 for a discussion.}
\ba
&&-i\tomega u_r-2\Omega u_\theta=-{\partial\over\partial r}(\eta+\phi),
\label{eq:fluid1}\\
&&-i\tomega u_\theta +{\kappa^2\over 2\Omega}u_r=-{im\over r}(\eta+\phi),\label{eq:fluid2}\\
&&-i\tomega u_z=-{\partial \over\partial z}(\eta+\phi),\label{eq:fluid3}\\
&&-i\tomega {\rho_0\over c^2}\eta+{1\over r}{\partial\over\partial
r} (r\rho_0 u_r)+{im\over r}\rho_0 u_\theta +{\partial\over\partial
z}(\rho_0 u_z)=0. \label{eq:fluid}\ea 
Here $\kappa$ is the epicyclic
frequency, defined by \be \kappa^2={2\Omega\over r}{d\over
dr}(r^2\Omega). \ee 
In this paper we will consider cold, (Newtonian)
Keplerian disks, for which the three characteristic frequencies,
$\Omega,\Omega_\perp$ and $\kappa$, are identical and equal to the
Keplerian frequency $\Omega_K=(GM/r^3)^{1/2}$. 
However, we continue to use different notations ($\Omega,\Omega_\perp,
\kappa$) for them in our treatment below when possible so that 
the physical origins of various terms are clear.

Following Tanaka et al.~(2002) and Takeuchi \& Miyama (1998) (see also
Okazaki et al.~1987; Kato 2001), 
we expand the perturbations with Hermite polynomials $H_n$:
\ba
\left[\begin{array}{c}
\phi(r,z)\\
\eta(r,z)\\
u_r(r,z)\\
u_\theta(r,z)\end{array}\right]
&=& \sum_n \left[\begin{array}{c}
\phi_n(r)\\
\eta_n(r)\\
u_{rn}(r)\\
u_{\theta n}(r)\end{array}\right] H_n(Z),\nonumber\\
u_z(r,z)&=& \sum_n u_{zn}(r)H_n'(Z), \label{eq:expand}\ea 
where $H'_n=dH_n/dZ$. The Hermite polynomials $H_n(Z)\equiv(-1)^n
e^{Z^2/2}d^n(e^{-Z^2/2})/dZ^n$ satisfy the following equations:
\ba
&& H_n''-ZH_n'+nH_n=0,\\
&& H_n'=nH_{n-1},\\
&& ZH_n=H_{n+1}+nH_{n-1},\\
&& \int_{-\infty}^{\infty}\!\exp(-Z^2/2)H_n H_{n'}dZ=\sqrt{2\pi}
\,n!\,\delta_{nn'}.\label{n!}
\ea
We note that any complete set of functions can be used as the basis of
the expansion. For an example, Lubow(1981) used Taylor series in the
expansion of vertical variable. However, choosing the Hermite
polynomials as the basis set, as we shall see, greatly simplifies the
mathematics involved, since they are eigenmodes in
variable $z$ for locally isothermal disks with a constant scale
height $h$ and quasi-eigenmodes for disks with a small radial
variation of $h$. 
Note that since $H_1=Z$, $H_2=Z^2-1$, the $n=1$ mode 
coincides with the bending mode studied by Papaloizou \& Lin (1995) 
(who considered disks with no resonance), and the $n=2$ mode is similar 
to the mode studied by Lubow (1981).

With the expansion in (\ref{eq:expand}), the fluid equations (\ref{eq:fluid})
become
\ba &&-i\tomega u_{rn}-2\Omega u_{\theta n}=
-{d\over dr} w_n+{n\mu\over r}w_n+{(n+1)(n+2)\mu\over r}w_{n+2},\label{u1}\\
&&-i\tomega u_{\theta n} +{\kappa^2\over 2\Omega}u_{rn}=-{im\over r}w_n,\label{u2}\label{eq:fluida}\\
&&-i\tomega u_{zn}=-{w_n/h},\label{eq:fluidb}\\
&&-i\tomega {\eta_n\over c^2}+
\left({d\over dr}\ln r\sigma +{n\mu\over r}\right)u_{rn}+
{\mu\over r}u_{r,n-2}+{d\over dr} u_{rn}+{im\over r}u_{\theta n}
-{n\over h}u_{zn}=0,
\label{eq:fluid2}\ea
where 
\be
w_n\equiv \eta_n+\phi_n,
\ee
and
\be
\mu\equiv  {d\ln h\over d\ln r}.
\ee
Eliminating $u_{\theta n}$ and $u_{zn}$ from eqs.~(\ref{u1})-(\ref{eq:fluid2}),
we have
\ba
{dw_n\over dr}&=&{2m\Omega\over r\tomega}w_n-{D\over\tomega }iu_{rn}
+{\mu\over r}[nw_n+(n+1)(n+2)w_{n+2}],\label{eq:dwndr}\\
{du_{rn}\over dr}&=&-\left[{d\ln(r\sigma)\over dr}+
{m\kappa^2\over 2r\Omega\tomega}\right]u_{rn}+{1\over i\tomega}
\left({m^2\over r^2}+{n\over h^2}\right)w_n+{i\tomega\over c^2}\eta_n
-{\mu\over r}(n u_{rn}+u_{r,n-2}),\label{eq:durndr}
\ea
where we have defined
\be
D\equiv \kappa^2-\tomega^2=\kappa^2-(\omega-m\Omega)^2.
\ee
Finally, we eliminate $u_{rn}$ to obtain a single second-order 
differential equation for $\eta_n$ [see eq.~(21) of Tanaka et al.~2002]
\footnote{On the left-hand-side of eq.~(\ref{eq:main}), we have 
dropped the term $-r^{-1}(d\mu/dr)\left[nw_n+(n+1)(n+2)w_{n+2}
\right]$. This term does not change the results of this paper. Also,
for a Keplerian disk with $c=$constant, $h\propto r^{3/2}$ and
$d\mu/dr=0$. See \S 8 for a discussion.}:
\ba
&& \left[{d^2\over dr^2}+\left({d\over dr}\ln{r\sigma\over D}\right){d\over dr}
-{2m\Omega\over r\tomega}\left({d\over dr}\ln{\Omega\sigma\over D}\right)
-{m^2\over r^2}-{D(\tomega^2-n\Omega_\perp^2)\over c^2\tomega^2}
\right] w_n \nonumber\\
&&\quad +{\mu\over r}\Biggl[
\left({d\over dr}-{2m\Omega\over r\tomega}\right) w_{n-2}
+n\left({d\over dr}\ln{D\over\sigma}-{4m\Omega\over r\tomega}\right)w_n
\nonumber\\
&&\qquad\quad -(n+1)(n+2)\left({d\over dr}
-{d\over dr}\ln{D\over\sigma}+{2m\Omega\over
r\tomega}\right)w_{n+2}\Biggr] \nonumber\\
&&\quad -{\mu^2\over
r^2}\Bigl[(n-2)w_{n-2}+n(2n-1)w_n+n(n+1)(n+2)w_{n+2} \Bigr]=-{D\over
c^2}\phi_n. \label{eq:main}\ea

Obviously, for $\mu=0$, different $n$-modes are decoupled.
But even for $\mu\neq 0$, approximate mode separation can still be 
achieved: When we focus on a particular
$n$-mode, its excitation at the resonance is decoupled from the other
$n$-modes (see Sects.~4-6), provided 
that the orders of magnitudes of $\eta_{n\pm 2}$
and their derivatives are not much larger than $\eta_n$ and
$d\eta_n/dr$ --- we shall adopt this assumption in remainder of
this paper. Note that if $\eta_{n\pm 2}$ is much larger than $\eta_n$, the 
coupling terms must be kept. In this case, the problem can be
treated in a sequential way. After arranging the potential $\phi_n$'s 
in the order of their magnitudes, from large to small, we first
treat the mode with the largest $\phi_n$; in solving this ``dominant''
mode, the coupling terms involving the other ``secondary''
$n$-modes can be neglected. We then proceed to solve
the next mode in the sequence: Here the coupling terms with 
the dominant mode may not be neglected, but since the dominant mode
is already known, these coupling terms simply act as a ``source''
for the secondary mode and the torque formulae derived in the following 
sections (\S\S 4-7) can be easily modified to account for the
additional source.

In the absence of self-gravity, waves excited by external potential carry
angular momentum by advective transport. The time averaged transfer rate 
of the $z$-component of angular momentum across a cylinder of radius $r$
is given by (see Lynden-Bell \& Kalnajs 1972; GT;
Tanaka et al.~2002)
\be
F(r)=\Bigl\langle r^2 \int_{-\infty}^\infty\! dz\int_0^{2\pi}\! d\theta\,
\rho_0(r,z)
u_r(r,\theta,z,t)u_\theta(r,\theta,z,t)\Bigr\rangle.
\ee
Note that a positive (negative) $F(r)$ implies angular momentum 
transfer outward (inward) in radius.
Using $u_r(r,\theta,z,t)=\rmRe\,[{u_r}_{n}H_n(Z) e^{i(m\theta-\omega  t)}]$,
$u_\theta(r,\theta,z,t)=\rmRe\,[{u_\theta}_{n}H_n(Z)
e^{i(m\theta-\omega  t)}]$, and eqs.~(\ref{eq:rho0}) and (\ref{n!}), 
we find that the angular momentum flux associated with the $(n,m)$-mode is 
\be
F_n(r)=n!\,\pi r^2\sigma 
\,\rmRe \,(u_{rn}u_{\theta n}^*)
\ee
(recall that we do not explicitly write out the dependence on $m$).
Using eqs.~(\ref{u1}) and (\ref{u2}), this reduces to
(see Tanaka et al.~2002)
\be
F_n(r)={n!\,\pi m r\sigma\over D}\rmIm\left[
w_n{dw_n^*\over dr}-(n+1)(n+2){\mu\over r}w_nw_{n+2}^*\right],
\label{F0}\ee
where $w_n=\eta_n+\phi_n$.
To simplify eq.~(\ref{F0}) further, we shall carry out local 
averaging of $F_n(r)$ over a a scale much larger than the local
wavelength $|k|^{-1}$ (see GT). 
As we see in the next sections, the perturbation $\eta_n$
generated by the external potential $\phi_0$ generally consists of
a nonwave part $\bar\eta_n$ and a wave part $\tilde\eta_n$;
thus $w_n=\phi_n+\bar\eta_n+\tilde\eta_n$. 
The cross term between $(\phi_n+\bar\eta_n)$ and $\tilde\eta_n$ in 
eq.~(\ref{F0}) can be neglected after the averaging.
The coupling term ($\propto w_nw_{n+2}^*$) between different modes
can also be dropped because of the radial averaging and 
$|w_{n+2}|\lo |w_n|$ (see above).
Thus only the wave-like perturbation carries angular momentum,
and eq.~(\ref{F0}) simplifies to 
\be 
F_{n}(r)\approx
{n!\,\pi m r\sigma\over D}\rmIm\left(
\tilde\eta_n{d\tilde\eta_n^*\over dr}\right).
\label{F1}\ee
In \S\S 4-6, we will use eq.~(\ref{F0}) or (\ref{F1}) to calculate the 
angular momentum transfer by waves excited or dissipated 
at various resonances.

\section{Dispersion Relation and Resonances}

Before proceeding to study wave excitations, it is useful to
consider local free wave solution of the form
\be
\eta_n\propto \exp\left[i\int^r\!k(s) ds\right]~.
\ee
For $|kr|\gg 1$, from the eq.~(\ref{eq:main}), in the absence of the 
external potential, we find (see Okazaki et al.~1987; Kato 2001)
\be
(\tomega^2-\kappa^2)(\tomega^2-n\Omega_\perp^2)/\tomega^2=k^2c^2,
\label{eq:disp}\ee
where we have used $h=c/\Omega_\perp\ll r$ (thin disk),
and $m,n\ll r/h$ --- we will be concerned with such $m,n$ throughout this
paper. Obviously, for $n=0$
we recover the standard density-wave dispersion relation for 2D
disks without self-gravity. In Appendix A, we discuss
general WKB solutions of eq.~(\ref{eq:main}).

At this point it is useful to consider the special resonant
locations in the disk. These can be
recognized by investigating the singular points and turning
points of eq.~(\ref{eq:main}) or by examining the characteristics of 
the dispersion relation (\ref{eq:disp}).
For $\omega>0$, the resonant radii are

(i) Lindblad resonances (LRs), where $D=0$ or $\tomega ^2=\kappa^2$,
including outer Lindblad resonance (OLR) at $\tomega=\kappa$ and inner
Lindblad resonance (ILR) at $\tomega=-\kappa$. 
The LRs are apparent singularities of eq.~(\ref{eq:main})
--- we can see this from (\ref{eq:dwndr}) and (\ref{eq:durndr}) that
the wave equations are well-behaved and all physical quantities
are finite at $D=0$. The LRs are turning points 
at which wave trains are reflected or generated.
Note that the ILR exists only for $m\geq 2$.  

(ii) Corotation resonance (CR), where $\tomega=0$. In general,
the CR is a (regular) singular point of eq.~(\ref{eq:main}),
except in the special case of $n=0$ and $d(\Omega\sigma/\kappa^2)/dr
=0$ at corotation. Some physical quantities (e.g., azimuthal velocity
perturbation) are divergent at corotation. Physically, this sigularity
signifies that a steady emission or absorption of wave action
may occur there. Note that no CR exists for $m=0$.

(iii) Vertical resonances (VRs), where $\tomega^2=n\Omega_\perp^2$
(with $n\geq1$), including outer vertical resonance (OVR) at
$\tomega=\sqrt{n}\Omega_\perp$ and inner vertical resonance (IVR) at
$\tomega=-\sqrt{n}\Omega_\perp$. The VRs are turning points of
eq.~(\ref{eq:main}). The IVR exists only for $m>\sqrt{n}$.  Note that
for Keplerian disks and $n=1$, the LR and VR are degenerate.

For $\omega<0$, a Lindblad resonance (LR) exists only for $m=0$, where
$\omega=-\kappa$, and a vertical resonance (VR) may exist for
$m<\sqrt{n}$, but there is no corotation resonance in the disk.  

From the dispersion relation we can identify the wave propagation
zones for $\omega>0$ (see Fig.~1 and Fig.~2): (1) For $n=0$, the wave
zone lies outside the OLR and ILR (i.e., $r>r_{OLR}$ and $r<r_{ILR}$);
(2) For $n\ge 2$, the wave zones lie between ILR and OLR
($r_{ILR}<r<r_{OLR}$) and outside the IVR (if it exists) and OVR
($r<r_{IVR}$ and $r>r_{OVR}$); (3) For $n=1$, waves can propagate
everywhere.

The group velocity of the waves is given by 
\be c_g\equiv {d\omega\over
dk}={kc^2\over \tomega \Bigl[
1-(\kappa/\tomega)^2(n\Omega_\perp^2/\tomega^2)\Bigr]}.
\label{gv}\ee
The relative sign of $c_g$ and and the phase velocity $c_p=\omega/k$
is important for our study of wave excitations in \S\S 4-7: 
For $\omega>0$, positive (negative) $c_g/c_p$ implies
that trailing waves ($k>0$) carry energy outward (inward), while leading
waves ($k<0$) carry energy inward (outward). The signs of $c_g/c_p$ for
different propagation regions are shown in Figs.~1-2.
Note that for $n=0$ and $n\ge 2$, $c_p\rightarrow\infty$ and
$c_g\rightarrow 0$ as $k\rightarrow 0$ at the Linblad/vertical
resonances. But for $n=1$, we have 
\be
c_g={c\tomega^2\over\tomega^2+\kappa^2}\,{\rm sgn}\,(-k\tomega D).
\ee 
Thus $|c_g|\rightarrow c/2$ at the Lindblad/vertical resonances
\footnote{Of course, eqs.~(\ref{eq:disp})-(\ref{gv})
are valid only away from the resonances, so the limiting values discussed
here refer to the asymptotic limits as the resonances are approached.}.

\begin{figure}
\vspace{-70pt}
\centerline{\epsfbox{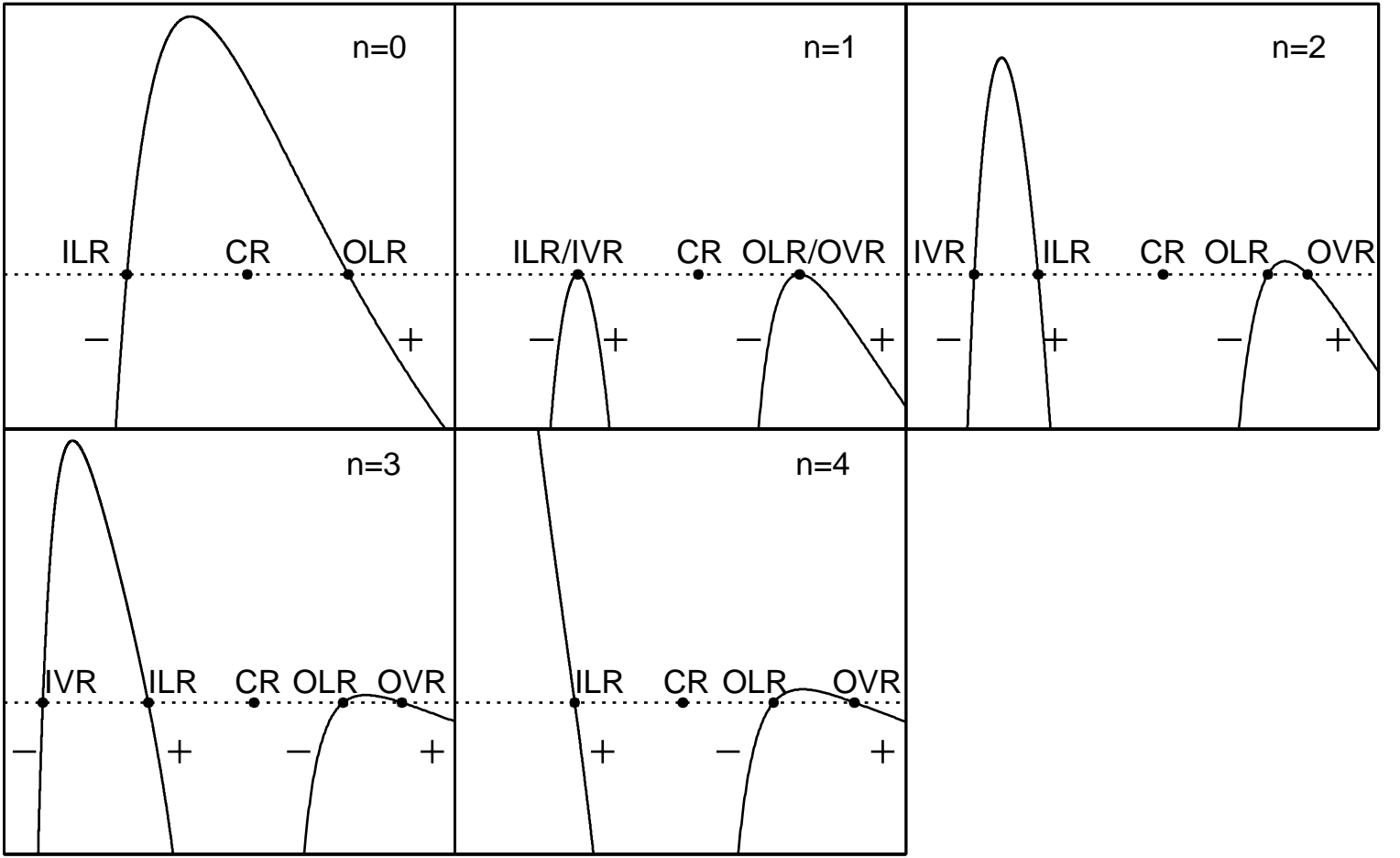}}
\vspace{-110pt}
\caption{A sketch of the function $G=D(1-n\Omega_\perp^2/\tomega^2)$
as a function of $r$ for $m=2$ and different values of $n$ (all for 
$\omega>0$). The dispersion relation is $G=-k^2c^2$, and thus 
waves propagate in the regions with $G<0$. The $\pm$ gives the sign 
of $c_g/c_p$ of waves in the propagation region.}
\end{figure}
\begin{figure} 
\vspace{-70pt}
\centerline{\epsfbox{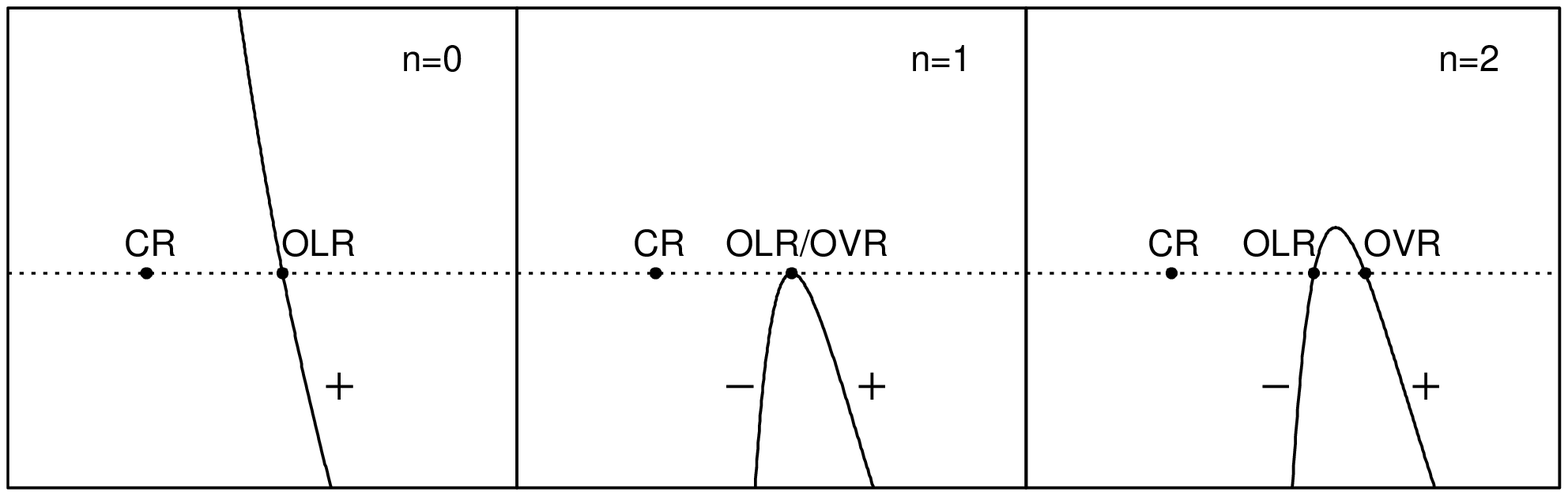}}
\vspace{-270pt}
\caption{Same as Fig.~1, except for $m=1$.}
\end{figure}

\section{Lindblad Resonances}

We now study wave excitations near a Lindblad resonance (where $D=0$).
Equation (\ref{eq:main}) shows that different $n$-modes
are generally coupled. However, when solving the equation 
for a given $n$-mode, the coupling terms can be neglected
if $|\eta_n|\ga |\eta_{n\pm 2}|$.
Note that in the vicinity of a Lindblad resonance,
$|d\ln D/dr|\gg |d\ln \sigma/dr|
\sim |d\ln \Omega/dr|\sim 1/r$. 
For a thin disk, if $m$ and $n$ not too large ($m,n\ll r/h$), 
the terms proportional to $c^{-2}\propto h^{-2}$
are the dominant non-singular terms. Keeping all the singular terms 
($\propto D^{-1}$), we have
\ba
&& \left[{d^2\over dr^2}-\left({d\ln D\over dr}\right){d\over dr}
+{2m\Omega\over r\tomega}\left({d\ln D\over dr}\right)
-{D(\tomega^2-n\Omega_\perp^2)\over c^2\tomega^2}
\right] \eta_n \nonumber\\
&&\quad +{\mu\over r}\left({d\ln D\over dr}\right)
\left[n\eta_n+(n+1)(n+2)\eta_{n+2}\right]
={d\ln D\over r dr}
\psi_n-{n\Omega_\perp^2 D\over c^2\tomega^2}\phi_n,
\label{eq:main1}\ea
where
\be
\psi_n\equiv \left(r{d\over dr}-{2m\Omega\over \tomega}-n\mu
\right)\phi_n-\mu (n+1)(n+2)\phi_{n+2}.
\label{eq:psin}\ee
Now the terms proportional to $(d\ln D/dr)\eta_n$
and $(d\ln D/dr)\eta_{n+2}$ can be dropped relative to 
the other terms (see Goldreich \& Tremaine 1978)\footnote
{To see this explicitly, we set $\eta_n=D^{1/2}y_n$ and reduce
eq.~(\ref{eq:main1}) to the form $d^2y_n/dr^2+f(r)y_n=\cdots$. 
We then find that the $(d\ln D/dr)\eta_n$ term in eq.~(\ref{eq:main1})
gives rise to a term $\propto (d\ln D/dr)\propto D^{-1}$ in $f(r)$, 
while the $d^2\eta_n/dr^2$ and $(d\ln D/dr)d\eta_n/dr$ terms in
eq.~(\ref{eq:main1}) both contribute terms 
proportional to $(d\ln D/dr)^2\propto D^{-2}$ in $f(r)$.},
we then obtain
\be
\left[{d^2\over dr^2}-\left({d\ln D\over dr}\right){d\over dr}
-{D(\tomega^2-n\Omega_\perp^2)\over c^2\tomega^2}
\right] \eta_n
={d\ln D\over r dr}\psi_n-{n\Omega_\perp^2 D\over c^2\tomega^2}\phi_n.
\label{eq:main2}\ee

We now proceed to solve eq.~(\ref{eq:main2}). Note that 
besides its (apparent) 
singular behavior, the resonance point of $D=0$ is a first-order
turning (or transition) point when $n\neq 1$, and a second-order
turning point when $n=1$ for Keplerian disks. These two cases should
be investigated separately.

\subsection{$n\neq 1$ Mode}

In the vicinity of the Lindblad resonance $r_L$, we change the independent
variable from $r$ to 
\be 
x\equiv (r-r_L)/r_L 
\ee 
and replace $D$ by
$(r\,dD/dr)_{r_L}x+(r^2\,d^2D/dr^2)_{r_L}x^2/2$. 
For $|x|\ll 1$, eq.~(\ref{eq:main2}) becomes 
\be
\left({d^2\over dx^2}-{1\over x}{d\over dx}-\beta x\right)\eta_n
={\psi_n\over x}-\alpha x-\gamma x^2 , \label{eq:main3}
\ee 
where $\psi_n$ on the right-hand-side is evaluated at $r=r_L$ and
\ba
&&\alpha={n r^3\phi_n dD/dr\over h^2\tomega^2}\Biggr|_{r_L},\quad
\beta={(\tomega^2-n\Omega_\perp^2)r^3dD/dr\over
h^2\Omega_\perp^2\tomega^2}\Biggr|_{r_L}, \quad\nonumber\\
&&\gamma=nr_L^4\left[{\phi_{n}d^2D/dr^2\over 2h^2\tomega^2}+{dD\over
dr}{d\over dr}\left({\phi_{n}\over h^2\tomega^2}\right)\right]_{r_L}.
 \label{l1}\ea
Here we have kept the term $-\gamma x^2$ in the Taylor expansion of the
second term of right-hand side of Eq.(\ref{eq:main2})
because the leading order term $-\alpha x$ generates only non-wave
oscillations, as we shall see later. In orders of magnitudes,
$|\beta|\sim (r_L/h)^2$ and $|\alpha|\sim |\gamma|\sim n(r_L/h)^2 |\phi_n|$.
In particular, for disks with $\Omega_\perp=\kappa=\Omega$ we have
\be
\beta=2(1-n)(1\pm m)\left({r^2\over h^2}{d\ln\Omega\over d\ln r}
\right)_{r_L},
\label{eq:beta}\ee
where upper (lower) sign refers to the OLR (ILR). Note that ILR occurs 
only for $m\ge 2$, so the factor $(1\pm m)$ is never zero.

To solve eq.~(\ref{eq:main3}), it is convenient to introduce a new
variable $\hat{\eta}$ defined by $\eta_{n}=d\hat{\eta}/dx$ (see Ward
1986).  We find 
\be 
{d\over dx}\left({1\over x}{d^2\hat{\eta}\over
dx^2}\right)-\beta {d\hat{\eta}\over dx} ={\psi_{n}\over
x^2}-\alpha-\gamma x .  \ee
Integrating once gives
\be
{d^2\hat{\eta}\over dx^2}-\beta x\hat{\eta}
=-\psi_{n}-\alpha x^2-{1\over 2}\gamma x^3+c x,
\ee
where $c$ is an integration constant. Then, let
\be
y=\hat{\eta}-{\alpha\over\beta}x-{\gamma\over 2\beta}x^2+{c\over\beta},
\ee
we have
\be
{d^2y\over dx^2}-\beta xy
=-\psi_{n}-{\gamma \over \beta}\equiv \Psi_{n}.
\label{eq:y}
\ee
As we see below, introducing the variable $y$ 
singles out the wave part from $\eta_n$.  
The homogeneous form of Eq.(\ref{eq:y}) is the Airy equation and its 
two linearly independent solutions are Airy functions 
(Abramowiz \& Stegun 1964, p.~446): 
\be
y_1=Ai(\beta^{1/3}x),\quad 
y_2=Bi(\beta^{1/3}x).  
\ee 
By the method of variation of parameters, the general solution to
the inhomogeneous equation (\ref{eq:y}) can be written as 
\be
y=y_2\int^x_0 y_1{\Psi_{n}\over W}dx-y_1\int^x_0 y_2{\Psi_{n}\over
W}dx+My_1+Ny_2, 
\label{eq:ygeneral}\ee 
where $M$ and $N$ are constants which should
be determined by the boundary conditions,
and $W=y_1dy_2/dx-y_2dy_1/dx=\beta^{1/3}/\pi$ is the Wronskian.
After writing $\xi=\beta^{1/3}x$, eq.~(\ref{eq:ygeneral}) becomes
\be
y=
{\pi\Psi_{n}\over\beta^{2/3}}
\left[\left(\int^\xi_0Ai(\xi)d\xi+N\right)Bi(\xi)
-\left(\int^\xi_0 Bi(\xi)d\xi+M\right)Ai(\xi)\right],
\label{eq:sol}
\ee
where we have absorbed a factor $\beta^{2/3}/(\pi\Psi_n)$ into
the the constants $M$ and $N$.

The constants $M,~N$ in eq.~(\ref{eq:sol}) are determined by boundary
conditions at $|\xi|\gg 1$. Note that the condition
$|x|=|\beta^{-1/3}\xi|\ll 1$, which is the regime of validity for the
solution described in this subsection, implies that $|\xi|\ll
|\beta|^{1/3}$. Since $|\beta|\gg 1$, it is reasonable to consider the
$|\xi|\gg 1$ asymptotics of eq.~(\ref{eq:sol}). In the following we
sometimes denote this asymptotics as $|\xi|\rightarrow\infty$ for
convenience.

For $\xi\gg1$, the asymptotic expressions of the Airy functions and their 
integrals are (Abramowiz \& Stegun 1964)
\ba
 &&Ai(\xi)\approx {1\over2}\pi^{-{1/2}}\xi^{-{1/4}}e^{-{2\over3}\xi^{3/2}}
\rightarrow 0,\quad
\int_0^\xi Ai(\xi)d\xi\approx {1\over3}-{1\over2}\pi^{-{1/2}}\xi^{-{3/4}}
e^{-{2\over3}\xi^{3/2}}\rightarrow{1\over3},\nonumber\\
&&Bi(\xi)\approx \pi^{-{1/2}}\xi^{-{1/4}}e^{{2\over3}\xi^{3/2}}
\rightarrow\infty,\quad
\int_0^\xi Bi(\xi)d\xi\approx \pi^{-{1/2}}\xi^{-{3/4}}e^{{2\over3}
\xi^{3/2}}\rightarrow\infty.
 \ea
Since $Bi(\xi)$ grows exponentially when $\xi\rightarrow\infty$, 
the coefficient $[\int^\xi_0 Ai(\xi)d\xi+N]$ before it in (\ref{eq:sol}) 
must be very small, otherwise the quantity $y$ (and hence $\eta_n$) will 
become exponentially large, making it impossible to match with any 
reasonable physical boundary conditions. Without loss of generality, we will
take this coefficient to be zero, based on the observation 
[see eq.~(\ref{airy-}) below]
that the solution on the $\xi<0$ side is nearly unaffected whether 
the coefficient is small or precisely zero. 
Thus $N =-\int^\infty_0 Ai(\xi)d\xi = -1/3$.
Note that although $\int_0^\xi Bi(\xi)d\xi$ in (\ref{eq:sol}) also 
exponentially grows when $\xi\rightarrow\infty$, it is canceled by
$Ai(\xi)$ which is exponentially small.
As the Airy functions are monotonic for $\xi>0$, wave solution
does not exist on the side of $\xi>0$, or 
\be
(1-n)(1\pm m)x<0 \qquad {\rm (Nonwave~ region)}
\ee
[see eq.~(\ref{eq:beta})]. 
This is consistent with the wave propagation diagram
discussed in \S 3 (see Figs.~1-2).

Now let us examine the $\xi<0$ region. For $\xi\ll-1$,
the asymptotic behaviors of the Airy functions are
\ba
&&Ai(\xi)\approx \pi^{-{1/2}}(-\xi)^{-{1/4}}\sin X(\xi),\nonumber\\
&&\int_0^\xi Ai(\xi)d\xi\approx -{2\over3}+\pi^{-{1/2}}
(-\xi)^{-{3/4}}\cos X(\xi),\nonumber\\
&&Bi(\xi)\approx \pi^{-{1/2}}(-\xi)^{-{1/4}}\cos X(\xi),\nonumber\\
&&\int_0^\xi Bi(\xi)d\xi\approx -\pi^{-{1/2}}(-\xi)^{-{3/4}}\sin X(\xi),
\label{airy-}\ea
where
\be
X(\xi)\equiv {2\over3}(-\xi)^{3/2}+{\pi\over4}.
\ee
Equation (\ref{eq:sol}) with $N=-1/3$ yields, for
$\xi\ll -1$, 
\be
y \rightarrow 
-{\pi^{1/2}\Psi_{n}\over2\beta^{2/3}}(-\xi)^{-{1/4}}
\left[(1-iM)e^{iX(\xi)}+ (1+iM)e^{-iX(\xi)}\right].
\ee
From the relation between $\eta_{n}$ and $y$, we then obtain the asymptotic 
expression for $\eta_{n}$ at $\xi\ll -1$ as
\be
\eta_{n}\rightarrow {\alpha\over\beta}+{\gamma\over\beta}x+
 i{\pi^{1/2}\Psi_{n}\over2\beta^{1/3}}(-\xi)^{{1/4}}
\left[(1-iM) e^{iX(\xi)}- (1+iM)e^{-iX(\xi)}\right].
\label{eta}
\ee
The first two terms in eq.~(\ref{eta}) describe the non-propagating 
oscillation, which is called non-wave part by GT.
The last term gives traveling waves. Eq.~(\ref{eta}) explicitly shows
that waves exist on the side of $\xi=\beta^{1/3}x<0$, or 
$(1-n)(1\pm m)x>0$. Again, this is consistent with the propagation diagram
discussed in \S 3: If $\omega>0$, the wave zones are located 
on the outer side of the OLR and inner side of the ILR for $n=0$, 
whereas for $n\ge 2$, waves exist on the inner side of the OLR and
outer side of the ILR. If $\omega<0$, LR is possible only for $m=0$, 
and the wave zone lies in the outer side of the resonance for $n=0$ 
and appears on the inner side for $n\geq 2$.

To determine the constant $M$, we require that waves excited by 
the external potential propagate away from the resonance.
The direction of propagation of a real wave is specified by its 
group velocity, as given by eq.~(\ref{gv}). 
For the waves going away from the resonance, we require
${\rm sgn}[c_g]={\rm sgn}[x]$, and thus the local wave-number $k$ must satisfy
${\rm sgn}[k]={\rm sgn}[x\tomega (1-n)]$.
On the other hand, the wavenumber associated with the 
wave term $e^{\pm iX(\xi)}$ in eq.~(\ref{eta}) is 
$k\equiv \pm dX/dr=\mp\beta^{1/3}(-\xi)^{1/2}r_L^{-1}$, which gives
${\rm sgn}[k]=\mp {\rm sgn}[\beta]=\pm {\rm sgn}[x]$ (since $\xi<0$).
Accordingly, if $\omega>0$, for the $n=0$ OLR and the $n\geq2$ ILR,
the $e^{iX}$ term represents the outgoing wave, and in order 
for the $e^{-iX}$ term to vanish, we have $M=i$, which leads to
\be
\eta_{n}\rightarrow \bar{{\eta}}+\tilde{\eta}=
{\alpha\over\beta}+{\gamma\over\beta}x+
i{\pi^{1/2}\Psi_{n}\over\beta^{1/3}}(-\xi)^{1/4}e^{iX(\xi)},
\quad (r>r_{_{OLR}}~{\rm for}~n=0~~{\rm or}~~
r>r_{_{ILR}}~{\rm for}~n\ge 2)
\label{eta1}
\ee
where $\bar{{\eta}}$ and $\tilde{\eta}$ represent the non-wave part and 
the wave part, respectively. Similarly, for
the $n=0$ ILR and the $n\geq2$ OLR, the $e^{-iX}$ term represents
the wave propagating away from the resonance, and eliminating the 
unwanted $e^{iX}$ term requires $M=-i$,
which yields
\be
\eta_{n}\rightarrow\bar{{\eta}}+\tilde{\eta}=
{\alpha\over\beta}+{\gamma\over\beta}x
- i{\pi^{1/2}\Psi_{n}\over\beta^{1/3}}(-\xi)^{1/4}e^{-iX(\xi)},
\quad (r<r_{_{ILR}}~{\rm for}~n=0~~{\rm or}~~
r<r_{_{OLR}}~{\rm for}~n\ge 2).
\label{eta2}
\ee
Lindblad resonance also occurs for $\omega<0$ and $m=0$,
in which case we find that eq.~(\ref{eta1}) applies for
$n\geq2$ and eq.~(\ref{eta2}) for $n=0$.

We can now use eq.~(\ref{F1}) to calculate the angular momentum flux
carried by the wave excited at the LR.  
Obviously, the $m=0$ mode carries no angular momentum.
For the $n=0$ OLR and the $n\geq2$ ILR, we substitute eq.~(\ref{eta1})
in eq.~(\ref{F1}), and find
\be
F_{n}(r>r_{_{LR}})
=-n!\,m\pi^2 \left({\sigma \Psi_{n}^2\over r dD/dr}\right)_{r_L},
\label{eq:flind1}\ee
where, according to eqs.~(\ref{eq:psin}), (\ref{l1}) and (\ref{eq:y}),
\ba
&& \Psi_{n} =\Biggl\{-r{d\phi_n\over dr}+{2m\Omega\over
\omega-m\Omega}\phi_{n}+n\left[\mu -{\Omega_\perp^2 r\over
2(\kappa^2-n\Omega_\perp^2)}{d\over dr}\ln\left({\phi_{n}^2\over
h^4\kappa^4}{dD \over dr}\right) \right]\phi_{n}\nonumber\\ 
&&\qquad +(n+1)(n+2)\mu\,\phi_{n+2}\Biggr\}_{r_L}.
\ea
Similarly, for $n=0$ ILR and $n\geq2$ OLR, we 
use eq.~(\ref{eta2}) in eq.~(\ref{F1}) to find
\be
F_{n}(r<r_{_{LR}})=n!\,m\pi^2 \left({\sigma \Psi_{n}^2\over r dD/dr}
\right)_{r_L}.
\label{eq:flind2}\ee
Obviously, the angular momentum flux at the $\xi>0$ side
(where no wave can propagate) vanishes. Note that the torque on 
the disk through waves excited in $r>r_{_{LR}}$ is $F_n(r>r_{_{LR}})$,
while for waves excited in $r<r_{_{LR}}$ the torque is
$-F_n(r<r_{_{LR}})$. Since $dD/dr<0$ for OLR and $>0$ for ILR, 
we find that for both $n=0$ and $n\ge 2$, the total torque on the disk
through both IRL and OLR is 
\be 
T_n({\rm OLR~and~ILR})=|F_{_{ORL}}|-|F_{_{ILR}}|
=-n!\,m\pi^2 
\left[\left({\sigma \Psi_{n}^2\over r dD/dr}
\right)_{\rm OLR}+\left({\sigma \Psi_{n}^2\over r dD/dr}
\right)_{\rm ILR}\right].
\label{eq:tlind}\ee
That is, independent of $n$, the torque on the disk is always 
positive at OLR and negative at ILR (see GT).
We note that for $n=\mu=0$, our result agrees with 
that for the 2D non-self-gravitating disks (GT;
Ward 1986).

\subsection{$n=1$ Mode: Lindblad/Vertical Resonances}

For $n=1$, the LR and VR are degenerate for a Keplerian disk,
and we shall call them Linblad/vertical resonances (L/VRs)
\footnote{Bate et al.~(2002) previously analysed the mixed
L/VRs for axisymmetric waves ($m=0$); such waves
do not carry angular momentum.}.
The resonance radius $r=r_L$ is both a (apparent) singular point and a 
second-order turning point of the wave equation (\ref{eq:main2}). 
The wave propagation diagram (see Fig.~1-2) shows that waves exist
on both sides of a L/VR. 

Expanding eq.~(\ref{eq:main2}) in the vicinity 
of the resonance ($|x|\ll 1$), we have
\be
{d^2\over dx^2}\eta_1-{1\over x}{d\over dx}\eta_1+b^2 x^2\eta_1
={\psi_{1}\over x}-\alpha_1 x , \label{eq:main4}
\ee
where
\be
\alpha_1={ r^3\phi_1 dD/dr\over h^2\tomega^2}\Biggr|_{r_L},\quad 
b=\Biggr|{r^2dD/dr\over h\Omega_\perp^2}\Biggr|_{r_L},\quad 
\psi_1=\left[\left(r{d\over dr}-{2m\Omega\over \tomega}-\mu
\right)\phi_1-6\mu \phi_{3}\right]_{r_L}.
\label{LVR-para}\ee
In orders of magnitudes, $|\alpha_1|\sim (r_L/h)^2\phi_1$ and
$b\sim r_L/h$. Substitution of $\eta_1=y-\psi_1x$ into 
eq.~(\ref{eq:main4}) gives
\be
{d^2\over dx^2}y-{1\over x}{d\over dx}y+b^2 x^2y
=b^2{\psi_{1}}x^3-\alpha_1 x . \label{eq:y1}
\ee
The two linearly independent solutions to the corresponding homogeneous 
form of eq.~(\ref{eq:y1}) are 
\be
y_1=e^{-ibx^2/2},\quad y_2=e^{ibx^2/2}.
\label{eq:y1y2}\ee
The method of variation of parameters then gives the general solution of
eq.~(\ref{eq:y1}):
\ba
y=e^{i\zeta^2/2}\left[\int^\zeta_{-\infty} e^{-i\zeta^2/2}S(\zeta)d\zeta
+N\right]
+e^{-i\zeta^2/2}\left[\int^{\infty}_\zeta e^{i\zeta^2/2}S(\zeta)d\zeta
+M\right],
\label{ynh}
\ea
where we have defined $\zeta=b^{1/2}x$ and 
\be 
S(\zeta)={\psi_1\zeta^2\over 2ib^{1/2}}-{\alpha_1\over 2ib^{3/2}}.
\ee
Note that although our analysis is limited to $|x|\ll 1$, we have extended
the integration limit to $\zeta=\pm\infty$ in eq.~(\ref{ynh}) ---
This is valid because $b\sim r_L/h\gg1$ for a thin disk and 
the integrands in the integrals are highly oscillatory for $|\zeta|\gg 1$
(so that the contribution to the integrals from the $|\zeta|\gg 1$ region 
is negligible; see Wong 2001). The wave solution 
in the $|\zeta|\gg 1$ region is approximately given by 
\ba
y&=&\left[\int^\infty_{-\infty} e^{-i\zeta^2/2}S(\zeta)d\zeta +N\right]
e^{i\zeta^2/2}+Me^{-i\zeta^2/2},\qquad (\zeta\gg 1)\label{ynh1}\\
y&=&\left[\int^{\infty}_{-\infty} e^{i\zeta^2/2}S(\zeta)d\zeta +M
\right]e^{-i\zeta^2/2}+Ne^{i\zeta^2/2},\qquad
(\zeta\ll -1).\label{ynh2}
\ea
The constants $M$ and $N$ can be fixed by requiring that no waves
propagate into the resonance.
From our analysis of the wave group velocity in \S 3 (see Figs.~1-2),
we find that for the waves propagating away from the resonance, the local 
wavenumber must be positive, irrespective of whether it is inner or outer
$n=1$ L/VR. Accordingly, we must have $M=0$ [from eq.~(\ref{ynh1})] and 
$N=0$ [from eq.~(\ref{ynh2})]. Since the integral
\be
\int^\infty_{-\infty} e^{\pm i\zeta^2/2}S(\eta)d\zeta
=\sqrt{\pi\over 2}\left(\pm {\psi_1\over b^{1/2}}+i{\alpha_1\over b^{3/2}}
\right)e^{\pm i{\pi/4}}.
\ee
eqs.~(\ref{ynh1})-(\ref{ynh2}) reduce to
 \ba
y &\simeq& \sqrt{\pi\over 2}
\left(-{\psi_1\over b^{1/2}}+i{\alpha_1\over b^{3/2}}\right)\,\,
e^{i{1\over2}\zeta^2-i{\pi\over4}},\qquad(\zeta\gg 1),\label{ynh10}\\
y &\simeq & \sqrt{\pi\over 2}
\left({\psi_1\over b^{1/2}}+i{\alpha_1\over b^{3/2}}\right)\,
e^{-i{1\over2}\zeta^2+i{\pi\over4}},\qquad (\zeta\ll -1).\label{ynh20}
\ea

Using eqs.~(\ref{F1}) and (\ref{ynh10}), we find that the angular 
momentum flux carried outward from the resonance (toward larger radii) 
by the wave excited at a L/VR is given by 
\be
F_{1}(r>r_{_{L/VR}})
=-{m\pi^2\sigma_L\over2r_L(dD/dr)_L}\left(\psi_1^2+{\alpha_1^2/b^2}\right)
\simeq -{m\pi^2\over 2}\left(
{r\sigma\phi_1^2\over h^2 dD/dr}\right)_{r_L},
\label{eq:f1>0}\ee
where in the second equality we have used the fact 
$|\alpha_1/b|=(r/h)|\phi_1|\gg |\psi_1|$.
Similarly, using (\ref{ynh20}), we find that the angular momentum flux 
carried by the $x<0$ wave is 
\be
F_1(r<r_{_{L/VR}})={m\pi^2\over 2}\left(
{r\sigma\phi_1^2\over h^2 dD/dr}\right)_{r_L}.
\label{eq:f1<0}\ee
Thus the angular momentum transfer to the disk through a
L/VR is simply $T_1({\rm L/VR})=F_{1}(r>r_{_{L/VR}})-F_1(r<r_{_{L/VR}})
=2F_{1}(r>r_{_{L/VR}})$.
Combining the inner and outer resonances, the total
torque is
\be
T_1({\rm OL/VR~and~IL/VR})=-m\pi^2 
\left[\left({r\sigma \phi_1^2\over h^2 dD/dr}
\right)_{\rm OL/VR}+\left({r\sigma \phi_1^2\over h^2 dD/dr}
\right)_{\rm IL/VR}\right].
\label{eq:tlind1}\ee
Again, we find that the torque on the disk is positive at OL/VR 
and negative at IL/VR.
Comparing the above result with the results of \S4.1 and \S 5, we find
that although the $n=1$ L/VR involves a combination of LR and VR
its behavior is more like a VR.

\section{Vertical Resonances ($n\ge 2$)}

We have already studied the $n=1$ vertical resonance in \S 4.2. 
We now examine VRs, $\tomega^2=n\Omega_\perp^2$, for $n\ge 2$.

In the neighborhood of a VR, there is no singularity in 
eq.~(\ref{eq:main}). For a thin disk ($h\ll r$) with $m,n\ll 
(r/h)$, it is only necessary to keep $d^2\eta_n/dr^2$ and 
the terms that are $\propto c^{-2}\propto h^{-2}$.
This can be justified rigorously from the asymptotic theory 
of differential equations (e.g., Olver 1974), which shows
that the discarded terms have no contribution to the leading order 
of the asymptotic solution. Indeed, the discarded 
terms only make a small shift to the vertical resonance for a thin disk.
As for the coupling terms with other modes, in addition 
to the reasons given in \S 4, the dropping of the coupling terms 
are more strongly justified here as the VRs with different $n$'s 
are located at different radii and their mutual effects can be 
considered insignificant. Therefore, around the VR radius $r_{_V}$,
we can simplify eq.~(\ref{eq:main}) to
\be 
{d^2\over dr^2}\eta_n
-{D(\tomega^2-n\Omega_\perp^2)\over h^2\tomega^2\Omega_\perp^2}
\eta_n =-{nD\over h^2\tomega^2}\phi_n.
\label{eq:a2}\ee
Changing the variable from $r$ to $x=(r-r_{_V})/r_{_V}$, we obtain,
for $|x|\ll 1$,  
\be
{d^2\over dx^2}\eta_n - \lambda x\eta_n =\chi, \label{eq:main30}
\ee
where
\be
\lambda=-{2r^3D(m\tomega d\Omega/dr+n\Omega_\perp
d\Omega_\perp/dr)\over
h^2\tomega^2\Omega_\perp^2}\Biggr|_{r_{_V}},\quad
\chi=-{nr^2D\phi_n\over h^2\tomega^2}\Biggr|_{r_{_V}}.  
\ee
Similar to \S 4.1, the general solution to eq.~(\ref{eq:main30}) reads
\be
\eta_n={\pi \chi\over \lambda^{2/3}}\left[
\left(\int^\varsigma_0 Ai(\varsigma)d\varsigma+N\right)Bi(\varsigma)
-\left(\int^\varsigma_0 Bi(\varsigma)d\varsigma+M\right)Ai(\varsigma)
\right],
\label{eq:sol3}
\ee
where $\varsigma=\lambda^{1\over 3}x$. Suppressing the mode which is
exponentially large when $\varsigma\rightarrow\infty$ yields
$N=-1/3$.

Waves can propagate in the $\varsigma<0$ region, i.e.,  
on the outer side of the OVR and on the inner side of the
IVR (see Figs.~1-2). For the waves to 
propagate away from the resonance, we require the group 
velocity to satisfy ${\rm sgn}[c_g]={\rm sgn}[x]$, and hence the local 
wavenumber to satisfy ${\rm sgn}[k]={\rm sgn}[x\tomega (1-n^{-1})]$.
Thus, if $\omega>0$, we demand that as $\varsigma\rightarrow-\infty$,
$\eta_n\propto e^{i{2\over3}(-\varsigma)^{3/2}}$ for the OVR
and $\eta_n\propto e^{-i{2\over3}(-\varsigma)^{3/2}}$ for the IVR.
This determines $M$ and yields, for $\varsigma\rightarrow -\infty$,
\ba
&&\eta_n\rightarrow 
-{\pi^{1/2}\chi\over \lambda^{2/3}}(-\varsigma)^{-{1\over 4}}\,
\exp\left[{i({2\over3}(-\varsigma)^{3/2}+{\pi\over4})}\right],\qquad
(r>r_{_{OVR}})\label{eq:ver1}\\
&&\eta_n\rightarrow 
-{\pi^{1/2}\chi\over \lambda^{2/3}}(-\varsigma)^{-{1\over 4}}\,
\exp\left[{-i({2\over3}(-\varsigma)^{3/2}+{\pi\over4})}\right],\qquad
(r<r_{_{IVR}}).\label{eq:ver2}
\ea
The angular momentum flux is then
\ba
&&F_{n}(r>r_{_{OVR}})
={n!\, m\pi^2}\left({\sigma\chi^2\over \lambda D}\right)_{r_{_V}}
=-{\pi^2\over2}n!\sqrt{n}{m\over\sqrt{n}+m}\left({r\sigma\phi_n^2\over h^2
\Omega d\Omega/dr}\right)_{\rm OVR},\label{eq:FOVR}\\
&&F_{n}(r<r_{_{IVR}})
=-{n!\, m\pi^2}\left({\sigma\chi^2\over \lambda D}\right)_{r_{_V}}
={\pi^2\over2}n!\sqrt{n}{m\over\sqrt{n}-m}\left({r\sigma\phi_n^2\over h^2
\Omega d\Omega/dr}\right)_{\rm IVR}.
\label{eq:FIVR}\ea
The torque on the disk is $F_{n}(r>r_{_{OVR}})$ at OVR and
$-F_{n}(r<r_{_{IVR}})$ at IVR. Note that the IVR exists only for $m>\sqrt{n}$
(see \S 3), so $-F_{n}(r<r_{_{IVR}})<0$. The total torque on the disk
due to both OVR and IVR is
\be
T_n({\rm OVR~and~IVR})=-{\pi^2\over2}n!\sqrt{n}
\left[{m\over\sqrt{n}+m}\left({r\sigma\phi_n^2\over h^2
\Omega d\Omega/dr}\right)_{\rm OVR}
\!\!+{m\over\sqrt{n}-m}\left({r\sigma\phi_n^2\over h^2
\Omega d\Omega/dr}\right)_{\rm IVR}\right].
\label{eq:tver}\ee
(Obviously, in the above expression, the IVR contribution should be set to
zero if $m<\sqrt{n}$). Again, we see that the torque is positive at OVR and
negative at IVR.

If $\omega<0$, a single VR exists for $m<\sqrt{n}$. 
In this case waves are generated on the outer side of the resonance with
\be
\eta_n(r>r_{_V})\rightarrow
-{\pi^{1/2}\chi\over \lambda^{2/3}}(-\varsigma)^{-{1\over 4}}
\exp\left[{-i({2\over3}(-\varsigma)^{3/2}+{\pi\over4})}\right],\qquad
(r>r_{_V};~~{\rm for}~\omega<0)
\label{eq:ver3}
\ee
as $\varsigma=\lambda^{1/3}x\rightarrow-\infty$. Thus
\be
F_{n}(r>r_{_V})={\pi^2\over2}n!\sqrt{n}{m\over\sqrt{n}- m}
\left({r\sigma\phi_n^2\over h^2\Omega d\Omega/dr}\right)_{r_{_V}},
\qquad (\omega<0).
\label{eq:FVR}
\ee
The torque on the disk due to such VR is negative.

It is interesting to compare our result with the one derived
from shearing sheet model (Takeuchi \& Miyama 1998; Tanaka et al 2002):
\be
F_{n}={\pi^2\over2}n!\sqrt{n}\left({r\sigma\phi_n^2\over h^2\Omega 
|d\Omega/dr|}
\right)_{r_{_V}}.
\label{shearing}\ee
Clearly, our eq.~(\ref{eq:FOVR}) reduces to eq.~(\ref{shearing}) 
for $m\gg\sqrt{n}$.

At this point, it is useful to compare the amplitudes of the waves
generated at different resonances.  
For LRs, $F_n\sim \sigma\phi_n^2/\Omega^2$;
for both the $n=1$ L/VRs and $n\ge 2$ VRs, $F_n\sim (r/h)^2\sigma
\phi_n^2/\Omega^2$.
Thus, when the external potential components have the same orders 
of magnitude, the angular momentum transfer through VRs 
is larger than LRs for thin disks.
Since we expect $\phi_n\propto (h/r)^n$, 
the $n=1$ vertical resonance may be comparable to
the $n=0$ Lindblad resonace in transferring angular momentum.

\section{Corotation Resonances}

Corotation resonances (CRs), where $\tomega=0$ or $\omega=m\Omega$,
may exist in disks for $\omega>0$. The WKB dispersion relation
and wave propagation diagram discussed in \S 3 (see Fig.~1-2)
show that for $n=0$, waves are evanescent in the region around 
the corotation radius $r_c$, while for $n>0$ wave propagation is 
possible around $r_c$. This qualitative difference is also reflected 
in the behavior of the singularity in eq.~(\ref{eq:main}).
We treat the $n=0$ and $n>0$ cases separately.

\subsection{$n=0$ Mode}

In the vicinity of corotation, we only need to keep the terms in
eq.~(\ref{eq:main}) that contain the $\tomega^{-1}$ singularity 
and the term $\propto h^{-2}$. The term $\propto \eta_{n+2}/\tomega$ can
also be dropped, since from eq.~(\ref{eq:dwndr}) we can see that the 
coupling term is negligible when $|\tomega|$ is small.
Thus for $n=0$, eq.~(\ref{eq:main}) can be approximately simplified to
\be
{d^2\over dr^2}\eta_0-{2m\Omega\over r\tomega} \left({d\over
dr}\ln{\sigma\Omega\over D}\right)\eta_0 -{D\over h^2\Omega_\perp^2}
\eta_0={2m\Omega\over r\tomega} \left({d\over dr}\ln{\sigma\Omega\over
D}\right)\phi_0+{4\mu m\Omega\over r^2\tomega}\phi_{2}.
\label{eq:a4}\ee
Near the corotation radius $r_c$, we introduce the variable 
$x=(r-r_c)/r_c$, and eq.~(\ref{eq:a4}) becomes 
\be
{d^2\eta_0\over dx^2}+\left({p\over x+i\epsilon}-q^2\right)
\eta_0=-{p\over x+i\epsilon}\Phi,
\label{c01}\ee
where
\ba
&&p=\left[{2\Omega\over d\Omega/dr}{d\over dr}\ln\left({\sigma\Omega\over D}\right)\right]_{r_c},\quad
q=\biggl|{Dr^2\over h^2\Omega_\perp^2}\biggr|^{1/2}_{r_c}=(\kappa r/c)_{r_c}
\gg 1,\\
&&\Phi=\left[\phi_0+{2\mu\phi_2 \over rd\ln({\sigma\Omega/D})/dr}\right]_{r_c}.
\ea
In eq.~(\ref{c01}), the small imaginary part $i\epsilon$
(with $\epsilon>0$) in $1/x$ arises because we consider the response of the
disk to a slowly increasing perturbation (GT)
or because a real disk always has some dissipation which has not been
explicitly included in our analysis. 
If $\mu=0$ or $|\phi_2|\ll |\phi_0|$, then the eq.~(\ref{c01}) is the same as
the equation studied by GT for 2D disks.

Goldreich \& Tremaine (1979) solved eq.~(\ref{c01}) neglecting the
$(p/x)\eta_0$ term, and found that there is a net flux of angular momentum  
into the resonance, given by
\be
\Delta F_c=F(r_c-)-F(r_c+)={2\pi^2m}\left[{\Phi^2\over d\Omega/dr}{d\over dr}
\left({\sigma\Omega\over D}\right) \right]_{r_c}.
\label{gt}\ee
On the other hand, integrating eq.~(\ref{c01}) from $x=0-$ to $x=0+$
gives $(d\eta_0/dx)_{0+}-(d\eta_0/dx)_{0-}=i\pi p(\eta_0+\Phi)$, 
which, together with eq.~(\ref{F0}), yield (Tanaka et~al.~2002)
\be
\Delta F_c={2\pi^2m}\left[{|\eta_0+\Phi|^2\over d\Omega/dr}{d\over dr}
\left({\sigma\Omega\over D}\right)\right]_{r_c}.
\label{c011}\ee
Tanaka et al.~(2002) argued that neglecting $\eta_0$ in 
eq.~(\ref{c011}) may be not justified in a gaseous disk and
that the revised formula (\ref{c011}) fits their numerical result 
better than eq.~(\ref{gt}).

Although eq.~(\ref{c011}) is a precise result of eq.~(\ref{c01}), 
the presence of the unknown quantity $\eta_0(r_c)$ makes the 
expression not directly useful. Therefore, a more rigorous solution of 
eq.~(\ref{c01}) without dropping the $(p/x)\eta_0$ term
seems desirable. In the following we provide such a solution.

After changing variable $w\equiv \eta_0+\Phi$, eq.~(\ref{c01}) becomes
\be
{d^2w\over dx^2}+({p\over x}-q^2)w=-q^2\Phi.
\label{c010}\ee
To solve this non-homogeneous equation, we need firstly to find the 
solutions of the corresponding homogeneous equation
\be
{d^2w\over dx^2}+({p\over x}-q^2)w=0.
\label{c02}\ee
By introducing the variables
\be
w=xe^{qx}y,\quad s=-2qx,
\ee
eq.~(\ref{c02}) is transformed to Kummer's equation 
(Abramowitz \& Stegun 1964)
\be
s{d^2y\over ds^2}+(2-s){dy\over ds}-(1+{p\over 2q})y=0.
\ee
Its two independent solutions can be chosen as
\be
y_1=U(1+{p\over 2q},2,s),\quad y_2=e^{s}U(1-{p\over 2q},2,-s),
\ee
where the function $U$ is a logarithmic solution defined as
\ba
&& U(a,n+1,z)={(-1)^{n+1}\over n!\Gamma(a-n)}\biggl[M(a,n+1,z)\ln z\nonumber\\
&&\qquad +\sum_{r=0}^\infty{(a)_rz^r\over(n+1)_rr!}
\{\psi(a+r)-\psi(1+r)-\psi(1+n+r)\}\biggr]\nonumber\\
&&\qquad +{(n-1)!\over \Gamma(a)}z^{-n}\sum^{n-1}_{r=0}
{(a-n)_rz^r\over(1-n)_rr!},
\label{u}\ea
in which $M(a,b,z)$ is Kummer's function, $\psi$ is the Digamma function 
and $(...)_r$ denotes the Pochhammer symbol (see Abramowitz \& Stegun 1964).
So the two independent solutions to eq.~(\ref{c02}) are
\be
w_1=xe^{qx}U(1+{p\over 2q},2,-2qx),\quad w_2=xe^{-qx}U(1-{p\over 2q},2,2qx),
\ee
and their Wronskian can be shown to be
\be
W=w_1{d\over dx}w_2-w_2{d\over dx}w_1=-{1\over 2q}e^{i\pi(1-{p\over2q})}.
\ee
Note that owing to the singularity of the function $U$ at the branch point 
$x=0$, a branch cut has been chosen in the lower half of the complex plane.

The solution to eq.~(\ref{c010}) has the form
\ba
w&=&w_2\int^x_{-\infty} w_1{-q^2\Phi\over W}dx+w_1\int^{\infty}_x
w_2{-q^2\Phi\over W}dx\nonumber\\ &=&2q^3\Phi
e^{-i\pi(1-{p\over2q})}xe^{-qx}U(1-{p\over
2q},2,2qx)\int^x_{-\infty}xe^{qx}U(1+{p\over 2q},2,-2qx)dx\nonumber\\
&&+2q^3\Phi e^{-i\pi(1-{p\over2q})}xe^{qx}U(1+{p\over
2q},2,-2qx)\int^{\infty}_xxe^{-qx}U(1-{p\over 2q},2,2qx)dx,
\ea
where we have adopted the physical boundary conditions that for
$|qx|\gg 1$, the quantity $|w|$ does not become exponentially large. 
One of the limits of each integration has been extend to infinity
since it only introduces negligible error due to the property of the 
integrand. From eq.~(\ref{u}), at $x=0$, we have
\ba
&&w(x=0)=-{\Phi\over 4}e^{-i\pi(1-{p\over2q})}{1\over \Gamma(1-{p\over 2q})}
\int^{\infty}_0e^{-{1\over2}t}t U(1+{p\over 2q},2,t)dt\nonumber\\
&&\qquad -{\Phi\over 4}e^{-i\pi(1-{p\over2q})}{1\over \Gamma(1+{p\over 2q})}\int^{\infty}_0e^{-{1\over2}t}t U(1-{p\over 2q},2,t)dt~~.
\ea
Utilizing a formula of Laplace transform (Erd\'{e}lyi et al.~1953)
\be
\int^\infty_0e^{-st}t^{b-1}U(a,c,t)dt={\Gamma(b)\Gamma(b-c+1)\over
\Gamma(a+b-c+1)}~{}_2F_1(b,b-c+1;a+b-c+1;1-s)
\ee
to calculate out the integrals, we obtain
\be
w(x=0)=-{\Phi\over 4}e^{-i\pi(1-{p\over2q})}{\sin{p\pi\over 2q}\over
{p\over 2q}\pi}\left[{{}_2F_1(2,1;2+{p\over2q};{1\over2})\over
1+{p\over2q}}+ {{}_2F_1(2,1;2-{p\over2q};{1\over2})\over
1-{p\over2q}}\right],
\ee
where ${{}_2F_1}$ denotes the Gaussian hypergeometric function. Hence,
\be
\eta_0(r_c)={1\over4}e^{i{p\pi\over2q}}{\sin{p\pi\over 2q}\over
{p\over 2q}\pi}\left[{{}_2F_1(2,1;2+{p\over2q};{1\over2})\over
1+{p\over2q}}+ {{}_2F_1(2,1;2-{p\over2q};{1\over2})\over
1-{p\over2q}}\right]{\Phi}-{\Phi},
\ee
and from eq.~(\ref{c011}),
\ba
&& \Delta F_c={2\pi^2m}\left[{\Phi^2\over d\Omega/dr}{d\over dr}
\left({\sigma\Omega\over D}\right)\right]_{r_c}\nonumber\\
&&\qquad\times{1\over16}{\sin^2{p\pi\over 2q}\over \left({p\over
2q}\pi\right)^2}\left[{{}_2F_1(2,1;2+{p\over2q};{1\over2})\over
1+{p\over2q}}+ {{}_2F_1(2,1;2-{p\over2q};{1\over2})\over
1-{p\over2q}}\right]^2.
\label{eq:tcor}\ea
By the fact ${}_2F_1(2,1;2;{1\over2})={2}$, it is easy to check 
that when $p/q\sim h/r \rightarrow 0$, $|\eta_0(r_c)|$ becomes
much smaller than $|\Phi|$, and $\Delta F_c$ reduces to eq.~(\ref{gt}), 
the original Goldreich-Tremaine result.

\subsection{$n\geq1$ Mode}

In the vicinity of corotation with $n\ge 1$,
the terms $\propto h^{-2}$ in eq.~(\ref{eq:main}) are dominant,
and we only need to keep these terms and the second-order differential term.
Eq.~(\ref{eq:main}) reduces to 
\be
{d^2\over dr^2}\eta_n
-{D(\tomega^2-n\Omega_\perp^2)\over h^2\tomega^2\Omega_\perp^2}\eta_n
=-{nD\over h^2\tomega^2}\phi_n,
\label{eq:a3}\ee
where $\phi_n$ is evaluated at $r=r_c$.
Defining $x=(r-r_c)/r_c$ and expanding eq.~(\ref{eq:a3}) around $x=0$, we have
\be
{d^2\eta_n\over dx^2}+C{1\over (x+i\epsilon)^2}\eta_n =-C{\phi_n\over 
(x+i\epsilon)^2}
\label{c1}\ee
where
\be
C={n\over m^2}\left({\kappa\over h\,d\Omega/dr}\right)^2_{r_c}\sim {n\over m^2}
\left({r_c\over h}\right)^2\gg 1.
\ee

We remark here that eq.~(\ref{eq:a3}) is similar to the equation derived 
in the context of the stability analysis of stratified flows 
(Booker \& Bretherton 1967), although the physics here is quite different.

The general solution to eq.~(\ref{c1}) is
\be
\eta_n=-\phi_n+Mz^{1/2}z^{i\nu}+Nz^{1/2}z^{-i\nu}
=-\phi_n+Mz^{1/2}e^{i\nu \ln z}+Nz^{1/2}e^{-i\nu \ln z},
\label{c10}\ee
where $\nu=\sqrt{C-{1\over4}}\gg 1$, $z=x+i\epsilon$ (with $\epsilon>0$)
and $M$ and $N$ are 
constants. The first term, the non-wave part, is a particular solution, 
while the other two terms are solutions to the homogeneous equation, 
depicting the waves.

From eq.~(\ref{gv}), we find that the group velocity of a wave (with wave 
number $k$) near corotation is given by $c_g=-c\tomega^2/
(\sqrt{n}\kappa\Omega_\perp) {\rm sgn}(k\tomega)$. 
Thus ${\rm sgn}(c_g)=-{\rm sgn}(kx)$ for $|x|>0$.
The $z^{1/2}z^{i\nu}$ component in eq.~(\ref{c10}) has local wave number 
$k=d(\nu\ln z)/dr=\nu/(r_c x)$, and thus it has group velocity $c_g<0$.
If we require no wave to propagate into $x=0$ from the $x>0$ region 
(as we did in studying LRs and VRs; see \S 4 and \S 5), 
then we must have $M=0$. 
Similarly, requiring no wave to propagate into corotation from the $x<0$ 
region gives $N=0$. Thus we have shown explicitly that 
waves are not excited at corotation. Further calculation shows that 
even adding the higher-order terms to eq.~(\ref{c1}) does not alter this 
conclusion. This is understandable because $|k|\rightarrow \infty$ 
near corotation, and short-wavelength perturbations couple weakly 
to the external potential.

However, unlike the $n=0$ case, waves with $n\ge 1$ 
can propagate into the corotation region and get absorbed
at the corotation. To calculate this absorption explicitly, let us consider 
an incident wave propagating from the $x>0$ region toward $x=0$:
\be 
\eta_n=A_+\, x^{1/2}e^{i\nu \ln x},\qquad (x>0;~~{\rm incident~wave}),
\label{eq:incident}\ee
where $A_+$ is constant specifying the wave amplitude.
To determine the transmitted wave, we note that $z=0$ is the branch
point of the function $e^{i\nu\ln z}$, and the physics demands
that the solution to eq.~(\ref{c1}) be analytic 
in the complex plane
above the real axis. Thus we must decide the branch cut of the
function so that it is single-valued. As discussed before (see \S 6.1),
while our analysis in this paper does not explicitly include dissipation, 
a real disk certainly will have some viscosity, and viscous effect can
be mimicked in our analysis by adding a small imaginary part $i\epsilon$
(with $\epsilon>0$) to the frequency $\omega$. Alternatively, 
this can be understood from causality where the disturbing 
potential is assumed to be turned on slowly in the distant past. 
This is the origin of imaginary part of $z=x+i\epsilon$ in eq.~(\ref{c1}) 
or eq.~(\ref{c10}). Therefore, we can choose the negative imaginary axis 
as the branch cut of the function $e^{i\nu\ln z}$. In doing so, 
we have $e^{i\nu\ln z}=e^{i\nu(\ln |x|+i\pi)}$ for $x<0$. Thus the transmitted
wave is given by
\be
\eta_n=i\, A_+\, e^{-\pi\nu} (-x)^{1/2}e^{i\nu\ln (-x)},\qquad
(x<0;~~{\rm transmitted~wave}).
\ee
Since $\nu\gg 1$, the wave amplitude is vastly decreased by a 
factor $e^{-\pi\nu}$ after propagating through the corotation. 
From eq.~(\ref{F0}) or (\ref{F1}), the net angular momentum flux absorbed
at corotation is 
\ba
&& \Delta F_c=F_n(r_c-)-F_n(r_c+)=
n!\,\pi m\left({\sigma\over \kappa^2}\right)_{r_c}\!\nu |A_+|^2(1+
e^{-2\pi\nu})\nonumber\\
&&\qquad \simeq n!\,\sqrt{n}\pi
\left({\sigma\over \kappa h |d\Omega/dr|}\right)_{r_c}|A_+|^2,
\label{deltafc}\ea
where in the last equality we have used $\nu\simeq \sqrt{C}\gg 1$.

Similarly, consider a wave propagating from $x<0$ toward $x=0$:
\be
\eta_n=A_- z^{1/2}e^{-i\nu\ln z}=i A_- e^{\pi\nu}(-x)^{1/2}e^{-i\nu\ln(-x)},
\qquad (x<0;~~{\rm incident~wave}).
\label{eq:incident2}\ee
The transmitted wave is simply
\be
\eta_n=A_-x^{1/2}e^{-i\nu\ln x},\qquad
(x>0;~~{\rm transmitted~wave}).
\ee
The net angular momentum flux into the corotation is 
\be
\Delta F_c=-n!\,\pi m\left({\sigma\over \kappa^2}\right)_{r_c}\!\nu 
|A_-e^{\pi\nu}|^2(1+e^{-2\pi\nu})
\simeq -n!\,\sqrt{n}\pi
\left({\sigma\over \kappa h |d\Omega/dr|}\right)_{r_c}|A_- e^{\pi\nu}|^2,
\label{deltafc1}\ee
where the negative value of $\Delta F_c$ arises because
waves inside corotation carry negative angular momentum.

In summary, for $n\ge 1$, waves propagating across the corotation are
attenuated (in amplitude) by a factor $e^{-\pi\nu}$. Thus the
corotation can be considered as a sink for waves with $n\ge 1$ 
--- a similar conclusion was reached by Kato (2003) and Li et al.~(2003), 
who were concerned with the stability of oscillation modes in accretion
disks around black hole.
The parameters $A_+,~A_-$ in eqs.~(\ref{eq:incident})
and (\ref{eq:incident2}) are determined by boundary conditions.  In
Appendix B, we discuss a specific example where waves excited at
Lindblad/vertical resonances propagate into the corotation and get
absorbed and transfer their angular momenta there.

\section{Wave Excitations at Disk Boundaries}

In the preceding sections we have examined the effects of various
resonances in a disk. Even without any resonance, density/bending waves
may be excited at disk boundaries.

To give a specific example, let us consider the $n=m=1$ bending wave in 
a Keplerian disk driven by a potential with frequency $\omega$ in the 
range $0<\omega<\Omega(r_{out})$. Here we use $r_{out}$ and
$r_{in}$ to denote the outer and inner radii of the disk. This situation 
occurs, for example, when we consider perturbations of the circumstellar
disk of the primary star driven by the secondary in a binary system.
Since no resonance condition is satisfied for the whole disk, 
the general solution to the disk perturbation is given by
[see eq.~(\ref{yq10}) in Appendix A]\footnote{The wavenumber $k$ for the
$n=m=1$ mode (for Keplerian disks) is given by $k^2h^2
=(\omega/\Omega)^2(2\Omega-\omega)^2/(\Omega-\omega)^2$, which
reduces to $k^2h^2=(2\omega/\Omega)^2\ll 1$ for $\omega<\Omega(r_{out})
\ll\Omega$. Thus the radial wavelength may be much larger than $h$ and 
comparable to $r$, in which case the WKB solution is not valid.}
\be
\eta_1=Q^{-1}G+(D/r\sigma)^{1/2}Q^{-1/4}
\left[M\exp(-i\int^r_{r_{in}}\!Q^{1/2}dr)+N\exp(i\int^r_{r_{in}}
\!Q^{1/2}dr)\right],
\label{yq10} \ee
where 
\be
Q=k^2={D(\Omega_\perp^2-\tomega^2)\over h^2\tomega^2\Omega_\perp^2},
\qquad 
G=-{D\over h^2\tomega^2}\phi_1.
\ee
To determine the constant $M,~N$, we assume that the inner boundary is 
non-reflective, 
thus $N=0$. For the outer boundary, we assume that the pressure
perturbation vanishes, i.e., $\eta_1=0$, this determines $M$ and 
eq.~(\ref{yq10}) then becomes
\be
\eta_1=Q^{-1}G-\left[{Q^{-1}G\over (D/r\sigma)^{1/2}Q^{-1/4}}\right]_{r_{out}}
\!\!(D/r\sigma)^{1/2}Q^{-1/4}\exp(i\int^{r_{out}}_r\!Q^{1/2}dr).
\ee
A direct calculation using eq.~(\ref{F1}) shows that the angular momentum 
flux carried by the wave is
\be
F_1=\pi\left({r\sigma G^2\over Q^{3/2}D}\right)_{r_{out}}
=\pi\left({r\sigma D\phi_1^2\over h^4\tomega^4Q^{3/2}}\right)_{r_{out}}.
\ee
Therefore, in this model, waves are mainly generated at the outer disk
boundary, propagating inward, while the angular momentum is transferred
outward.

\section{Discussion}

Here we discuss several assumptions/issues related to our theory
and how they might affect the results of the paper.

(i) {\it Radially Nonisothermal Disks.}
In deriving the basic equation (\ref{eq:main}) for the disk perturbation,
we have dropped several terms proportional to the radial gradient of 
the sound speed (see footnotes 1 and 2). It is easy to see that these
terms vary on the lengthscale of $r$ and do not introduce any 
singularity or turning point in our equation, therefore they do not affect
any of our results concerning wave excitation/damping studied in 
\S \S 4-7. However, a $r$-dependent sound speed gives rise to 
a nonzero $\partial \Omega/dz$, which can modify the structure of
the perturbation equation near corotation. Indeed, if the (unperturbed)
surface density $\sigma$ and sound speed $c$ profiles
satisfy simple power-laws, $\sigma\propto r^{-\alpha}$ and $c\propto 
r^{-\beta}$, then the angular velocity profile in the disk is 
given by (Tanaka et al.~2002)
\be
\Omega=\Omega_K\left\{1-{1\over2}\left({h\over
r}\right)^2\left[{3\over2}+\alpha+\beta\left({z^2\over
h^2}+1\right)\right]\right\},
\label{omegarz}\ee 
where $\Omega_K(r)$ is the Keplerian angular velocity.
For a thin disk, the deviation of $\Omega(r,z)$ from $\Omega_K$ is obviously
very small. Nevertheless, if the $z$-dependence of $\Omega$ is taken 
into account, an additional term,
$(-\beta m n \Omega D/r^2\tomega^3)w_n$ should be be added
to the left-hand-side of eq.~(\ref{eq:main})\footnote{Nonzero
$\partial\Omega/\partial z$ also gives rise to an additional ``coupling'' 
term, $\propto (\beta/\tomega^3)\eta_{n+2}$. This can be neglected 
[see the discussion following eq.~(\ref{eq:main})].}.
Obviously, this term does not affect waves near a Lindblad resonance and 
vertical resonance. Because of the strong $\tomega^{-3}$ singularity 
at corotation (where $\tomega=0$), one might suspect that our result 
on the $n\ge 1$ corotation resonance (see \S 6.2) may be affected 
(see Tanaka et al.~2002). In fact, we can show this is not the case. 
With the $\tomega^{-3}$ term included, the perturbation equation near 
a $n\ge 1$ corotation resonance is modified from eq.~(\ref{c1}) to
\be
{d^2\eta_n\over dx^2}+{C\over x^2}\eta_n+{C_1\over x^3}\eta_n 
=-{C\phi_n\over x^2}-{C_1\phi_n\over x^3},
\label{d1}\ee
where 
\be
C_1=-{n\beta\over m^2}\left[{\kappa^2\Omega\over r^3(d\Omega/dr)^3}
\right]_{r_c}.
\ee
Since $C\sim (n/m^2)(r_c/h)^2\gg 1$ while $C_1\sim n\beta/m^2$, the new
terms are important only for $|x|\lo \beta (h/r_c)^2$. Thus we expect that 
our solution, eq.~(\ref{c10}), remains valid for $|x|\gg \beta (h/r_c)^2$.
Indeed, the general solution of eq.~(\ref{d1}) is 
\be
\eta_n=-\phi_n+Mx^{1\over2}J_{i2\nu}(2C_1^{1\over2}x^{-{1\over2}})
+Nx^{1\over2}J_{-i2\nu}(2C_1^{1\over2}x^{-{1\over2}}),
\ee
where $\nu=\sqrt{C-1/4}$, and $J_{\pm 2i\nu}$ is the Bessel function
(Abramowitz \& Stegun 1964, p.~358). This solution approaches the form of 
eq.~(\ref{c10}) for $|x|\gg \beta (h/r_c)^2$.
Therefore, the analysis in \S 6.2 remains valid and our result on
wave absorption at the $n\ge 1$ corotation resonance is unchanged.
We conclude that while our theory is explicitly based on radially isothermal
disks, our results remain valid when this condition breaks down.

(ii) {\it Vertical structure of disks.}
Our theory is concerned with vertically isothermal disks, for which 
3-dimensional perturbations can be decomposed into various
Hermite components [eq.~(\ref{eq:expand}); see Tanaka et al.~2002]. 
It would be interesting to investigate if similar decomposition 
(with different basis functions) is possible for more general disk 
temperature profiles. To simplify our equations, we have also 
neglected stratification in our analysis [see eq.~(\ref{schwarz})].
In particular, vertical stratification gives rise to
a local Brunt-V\"ais\"al\"a frequency of order $\Omega_\perp (z/h)^{1/2}$,
and it is not clear to what extent such stratification will affect our results
involving vertical fluid motion in the disk (see Lubow \& Ogilvie 1998).
This issue requires further study.

(iii) {\it Non-Keplerian Disks.}  
Although in this paper we have considered Keplerian disks (for which
$\Omega=\kappa=\Omega_\perp$), extension to more general
disks is not difficult. Indeed, since we have been careful to
distinguish $\Omega,~\kappa,~\Omega_\perp$ throughout the paper,
most of our equations are valid when $\Omega\neq\kappa\neq\Omega_\perp$.
The only exception is the $n=1$ Lindblad/vertical resonances
studied in \S 4.2: Only for a Keplerian disk ($\Omega_\perp=\kappa$)
is the $n=1$ vertical resonance degenerate with the Lindblad resonance,
and such a combined Lindblad/vertical resonance needs special
treatment. For a disk where the Lindblad resonance and $n=1$
vertical resonance are well separated, they must be treated
separately, with the procedure similar to those given in
\S 4.1 (for Lindblad resonances) or \S 5 (for vertical resonances).
For a nearly Keplerian disk, with $|\Omega_\perp-\kappa|/\kappa \ll 1$,
the LR (at $r_L$) and the $n=1$ VR (at $r_V$) are rather close, 
and the problem requires some consideration.
Expanding eq.~(\ref{eq:main2}) around the Lindblad resonance, we find
[cf. eq.~(\ref{eq:main4})]
\be
{d^2\over dx^2}\eta_1-{1\over x}{d\over dx}\eta_1+b^2 x(x-x_V)\eta_1
={\psi_{1}\over x}-\alpha_1 x ,
\ee
where $x_V=(r_V-r_L)/r_L$ and $\alpha_1,~b,~\psi_1$ are given by 
eq.~(\ref{LVR-para}). Obviously, solution (\ref{eq:y1y2})
breaks down for $|x|\lo |x_V|$. For $bx_V^2\ll 1$, or $|x_V|\ll b^{-1/2}
\sim (h/r_L)^{1/2}$, the asymptotic solutions (\ref{ynh10})
and (\ref{ynh20}) remain valid and the angular momentum flux 
is unchanged. For $bx_V^2\go 1$, the angular momentum 
flux expression has the same form as eq.~(\ref{eq:f1>0})
or (\ref{eq:f1<0}) except that the pre-factor ($m\pi^2/2$) 
may be changed by a factor of order unity.

(iv) {\it Nonlinear effect.}  From the wave solutions we
have derived in this paper, we see that the enthalpy or density 
perturbation of the disk is finite at various resonances. 
However, the azimuthal 
velocity perturbation can become singular at 
the corotation resonance (GT).
Viscous and nonlinear effects may become important at
the resonance and affect the derived torque formula. 
Therefore, linear, inviscid theory for the 
corotation resonance is incomplete even for a very small
external potential. As has been pointed out by Balmforth \& Korycansky
(2001), and largely discussed in hydrodynamic stability
theory (see Drazin \& Reid 1981), critical layer may emerge at the
resonance. This issue requires further study
(see Ogilvie \& Lubow 2003; Goldreich \& Sari 2003).


\section{Conclusion}

In the paper we have studied the linear response of a 3D gaseous disk
to a rigidly rotating external potential. The disk is assumed to be
non-self-gravitating, geometrically thin and vertically isothermal. 
The external potential and the disk perturbation can be decomposed into 
various Fourier-Hermite components, each proportional to 
$H_n(z/h) \exp(im\theta-i\omega t)]$, characterized by the azimuthal 
index $m$ and the vertical index $n$ which specifies the number of nodes 
along the $z$-direction in the fluid density perturbation. We have 
derived analytical expressions for the various wave modes excited at 
Lindblad resonances and vertical resonances, and calculated the angular 
momentum fluxes associated with these waves and hence the torque on the 
disk from the external potential. We have also studied wave damping and 
angular momentum transfer at corotation resonances. For wave
modes that involves only 2D motion ($n=0$), our general formulae
reduce to the standard result of Goldreich \& Tremaine (1979).

Our main results on wave excitation/damping can be most 
conveniently summarized using the wave propagation diagram 
(Figs.~1-2) which follows from the dispersion 
relation [eq.~(\ref{eq:disp})]. In 2D disks, waves are excited
only at the inner and outer Lindblad resonances, and propagate
away from the corotation. By contrast, in 3D disks,
additional channels of wave generation open up through vertical
resonances, and waves can propagate into corotation and where 
angular momenta are deposited. Irrespective of the direction of
propagation of the excited waves, the torque on the disk is positive 
for waves generated at outer Linblad or vertical resonances and
negative at inner resonances.

Our paper contains a number of analytical results which are
obtained for the first time. A guide to the key equations are as follows:
(i) {\it Lindblad resonances:} For the $n=0$ and  $n\ge 2$ modes,
the wave amplitudes excited at the resonances are given by 
eqs.~(\ref{eta1})-(\ref{eta2}), the associated angular momentum 
fluxes eqs.~(\ref{eq:flind1}) and (\ref{eq:flind2}) and the torque
on the disk eq.~(\ref{eq:tlind}); for $n=1$, the Lindblad resonance
and vertical resonance coincide for a Keplerian disk, the corresponding
equations are (\ref{ynh10})-(\ref{ynh20}), (\ref{eq:f1>0})-(\ref{eq:f1<0})
and (\ref{eq:tlind1}).
(ii) {\it Vertical resonances:} For $n\ge 2$, the waves amplitudes
excited at the vertical resonances are given by 
eqs.~(\ref{eq:ver1})-(\ref{eq:ver2}), the angular momentum 
fluxes eqs.~(\ref{eq:FOVR})-(\ref{eq:FIVR}) and the torque
on the disk eq.~(\ref{eq:tver}).
(iii) {\it Corotation resonances:}
For $n=0$, waves cannot propagate around the corotation region, 
but a torque is delivered to the disk at corotation. An improved
expression for the torque is given by eq.~(\ref{eq:tcor}), which
reduces to the standard Goldreich-Tremaine result (\ref{gt})
in the $h/r\rightarrow 0$ limit. For $n>0$, waves can propagate into the
corotation. The angular momentum flux deposited to the disk
at corotation is given by eq.~(\ref{deltafc}) or (\ref{deltafc1}),
depending on the incident wave amplitude.

The last paragraph refers to waves excited by a prograde-rotating
potential (with pattern speed $\Omega_p=\omega/m>0$). 
It is of interest to note that for $m<\sqrt{n}$, vertical resonant
excitations exist for a retrograde-rotating potential
(with $\Omega_p<0$, i.e., the perturber rotates in the direction
opposite to the disk rotation): The excited wave has an amplitude
given by eq.~(\ref{eq:ver3}) and angular momentum flux given by
eq.~(\ref{eq:FVR}). 

Even without any resonance, waves can be excited at disk boundaries.
An example is discussed in \S 7.

An interesting finding of our paper is that for a given potential component
$\phi_n$ with $n\ge 1$, vertical resonances are much more efficient 
[by a factor of order $(r/h)^2$] than Lindblad resonances in transferring 
angular momentum to the disk. Whether vertical resonances can compete
with $n=0$ Lindblad resonances depend on the relative values of
$\phi_n$ ($n\ge 1$) and $\phi_0$. 
Since we expect $\phi_n\propto (h/r)^n$, the angular momentum transfer
through the $n=1$ vertical resonance may be as important as
the $n=0$ Lindblad resonace when considering the perturbation of a 
circumstellar disk by a planet in an inclined orbit.
We plan to investigate this and other related issues discussed 
in \S 8 in the future.

\appendix

\section{Solutions Away From Resonances}

In the region away from any of the resonances, we only need to retain 
in eq.~(\ref{eq:main}) the first and second derivative terms as well as
the terms $\propto h^{-2}$. Thus for $n>0$, eq.~(\ref{eq:main}) reduces to
\be
{d^2\over dr^2}\eta_n+\left({d\over dr}\ln{r\sigma\over D}\right)
{d\over dr}\eta_n+Q\eta_n=G,
\label{a5}\ee
where for $n>0$, 
\ba
&&Q=-{D(\tomega^2-n\Omega_\perp^2)\over h^2\tomega^2\Omega_\perp^2},\\
&&G=-{nD\over h^2\tomega^2}\phi_n,\qquad\qquad (n>0).
\ea
For $n=0$, all the potential terms in eq.~(\ref{eq:main}) must kept, thus
\ba
&&Q=-{D\over h^2\Omega_\perp^2},\\
&&G=\Biggl[-{d^2\over dr^2}-\left({d\over dr}\ln{r\sigma\over D}\right)
{d\over dr}+{2m\Omega\over r\tomega}\left({d\over dr}
\ln{\Omega\sigma\over D}\right)+{m^2\over r^2}\Biggr]\phi_0\nonumber\\
&&\qquad +{2\mu\over r} \left({d\over dr}-{d\over dr}
\ln{D\over\sigma}+{2m\Omega\over r\tomega}\right)\phi_{2},\qquad\qquad
(n=0).
\label{a6}\ea
Clearly, $Q=k^2$ gives the WKB dispersion relation (\ref{eq:disp}),
and $|Q|\sim h^{-2}$.
After making a transformation of the variable,
\be
y=(r\sigma/D)^{1/2}\eta_n,
\label{yq0}\ee
and using $|Q|\gg 1/r^2$, eq.~(\ref{a5}) reduces to 
\be 
{d^2y\over d^2r}+Qy=(r\sigma/D)^{1/2}G.
\label{nr}\ee

In the wave-propagating region ($Q>0$), the Liouville-Green or WKB 
approximation gives two independent solutions to the homogeneous 
equation of (\ref{nr}) (e.g., Olver 1974)
\be
y_1=Q^{-1/4}\exp(-i\int^r\!Q^{1/2}dr),\quad
y_2=Q^{-1/4}\exp(i\int^r\!Q^{1/2}dr).
\ee
A particular solution to eq.~(\ref{nr}) can be written asymptotically, 
owing to the large parameter $Q$, as $y=Q^{-1}(r\sigma/D)^{1/2}G$,
Thus the general solution to eq.~(\ref{a5}) is
\be
\eta_n=Q^{-1}G+(D/r\sigma)^{1/2}Q^{-1/4}
\left[M\exp(-i\int^r\!Q^{1/2}dr)+N\exp(i\int^r\!Q^{1/2}dr)\right],
\label{yq10} \ee
where the two constants $M$ and $N$ are determined by
boundary conditions. The boundary conditions may be imposed
at the true disk boundary, or by the requirement to match the wave coming
from a resonance.

The the evanescent region ($Q<0$), the general solution to eq.~(\ref{a5})
can be obtained in a similar way:
\be
\eta_n=Q^{-1}G+(D/r\sigma)^{1/2}|Q|^{-1/4}
\left[M\exp(-\int^r\!|Q|^{1/2}dr)+N\exp(\int^r\!|Q|^{1/2}dr)\right].
\label{yq10d} \ee

\section{An Example of Wave Excitation and Absorption}

Here we give an example of the global solution of wave excitation at a
Lindblad/vertical resonance, wave propagation, and wave absorption
at corotation. For definiteness, we consider the $n=1$ case,
and assume that the wave excited at the L/VR (see \S 4.2) does not suffer
any damping as it propagates toward the corotation radius.
The wave excited from the outer L/VR at $r=r_{_{OL}}$ is given by 
eq.~(\ref{ynh20}) and can be rewritten as
\be
\eta_1=-i\sqrt{\pi\over 2}\,e^{i\pi/4}\left({\Omega_{\perp}\phi_1\over h^{1/2}
|dD/dr|^{1/2}}\right)_{r_{_{OL}}}\!\!
\exp\left[-i{1\over 2}\biggl|{dD/dr\over h\Omega_{\perp}^2}
\biggr|_{r_{_{OL}}}\!(r-r_{_{OL}})^2\right].
\label{c5}
\ee
For the region between the corotation and the outer L/VR, $r_c<r<r_{_{OL}}$,
the general solution is given by eq.~(\ref{yq10}). Note that in this region,
the group velocity $c_g$ has opposite sign as the phase velocity (see 
Figs.~1-2). Keeping only the inward-propagating wave term, we have
\be
\eta_1=N(D/r\sigma)^{1/2}Q^{-1/4}\exp(i\int^r_{r_0}Q^{1/2}dr),
\label{c6}\ee
where $r_0$ is a fiducial radius ($r_c<r_0<r_{_{OL}}$).
Matching (\ref{c6}) with (\ref{c5}) as $r\rightarrow r_{_{OL}}$ gives
\be
N=-i\sqrt{\pi\over 2}\,e^{i\pi/4}
\left[\left({r\sigma\over |dD/dr|}\right)^{1/2}
\! h^{-1}\phi_1\right]_{r_{_{OL}}}\exp(-i\int_{r_0}^{r_{_{OL}}}Q^{1/2}dr).
\ee
Matching (\ref{c6}) with $\eta_1=A_+ x^{1/2}e^{i\nu \ln x}$ [see 
eq.~(\ref{eq:incident})] as $r\rightarrow r_c$ 
[or $x=(r-r_c)/r_c\rightarrow 0+$] yields
\be 
|A_+|=\sqrt{\pi m\over 2}
\left({h\Omega |d\Omega/dr|\over\sigma}\right)^{1/2}_{r_c}
\left[\left({r\sigma\over |dD/dr|}\right)^{1/2}
\! h^{-1}\phi_1\right]_{r_{_{OL}}}.
\ee
The angular momentum flux associated the wave at $r=r_c+$ is then
[see eq.~(\ref{deltafc})]
\be
F_1(r=r_c+)=-\pi\left({\sigma\over \Omega h |d\Omega/dr|}\right)_{r_c}|A_+|^2
=-{\pi^2m\over 2}\left({r\sigma\phi_1^2\over h^2|dD/dr|}\right)_{r_{_{OL}}}.
\ee
That is, the angular momentum absorbed at $r=r_c+$ is $-F_1(r=r_c+)$.
Not surprisingly, this is exactly equal to the inward angular momentum flux
transferred away from the OL/VR, i.e., $-F_1(r<r_{_{OL/VR}})$
[see eq.~(\ref{eq:f1<0})].

We can similarly show that the angular momentum transferred by the 
waves excited at the inner Lindblad resonance and propagating to 
the corotation is absorbed at corotation, as well as the waves themselves.



\section*{Acknowledgments}
We thank the referee for useful comments which improved this paper.
H.Z. thanks the hospitality of the Astronomy department at 
Cornell University, where this work was carried out. 
D.L. thanks IAS (Princeton), CITA (Toronto), TIARA (Taiwan),
KITP (Santa Barbara) and NAOC (Beijing) and Tsinghua University (Beijing) 
for extended visits during the period of this research.
This work was supported in part by the Fok Ying Dung Education Foundation,
by the NSF grant AST 0307252 and NASA grant NAG 5-12034.

\bsp
\label{lastpage}

\begin{thebibliography}{}

\bibitem[]{} 
Abramowitz, M., Stegun, I.A., 1964, Handbook of Mathematical 
Functions, Dover, New York

\bibitem[]{}
Artymowicz, P., 1993, ApJ,  419, 155

\bibitem[]{}
Artymowicz, P., 1994, ApJ, 423, 581

\bibitem[]{}
Balmforth, N.J., Korycansky, D.G., 2001, MNRAS, 326, 883

\bibitem[]{}
Bate, M.R., Ogilvie, G.I., Lubow, S.H., \& Pringle, J.E.~2002, MNRAS, 
332, 575

\bibitem[]{}
Booker, J. \& Bretherton, F.P., 1967, J. Fluid Mech., 27, 513

\bibitem[]{}
Drazin, P.G., Reid, W.H., 1981, Hydrodynamic Stability, Cambridge University Press, Cambridge

\bibitem[]{}
Erd\'{e}lyi, A., et al, 1953, Higher Transcendental Functions, Vol.1, McGraw-Hill, New York

\bibitem[]{}
Goldreich, P. \& Sari, R. 1993, ApJ, 585, 1024

\bibitem[]{}
Goldreich, P. \& Tremaine, S. 1978, Icarus, 34, 240

\bibitem[]{}
Goldreich, P. \& Tremaine, S. 1979, ApJ, 233, 857 (GT)

\bibitem[]{}
Goldreich, P. \& Tremaine, S. 1980, ApJ, 241, 425

\bibitem[]{}
Kato, S. 2001, PASJ, 53, 1

\bibitem[]{}
Kato, S. 2003, PASJ, 55, 257

\bibitem[]{}
Korycansky, D.G., Pollack, J.B.,1993, Icarus, 102, 150

\bibitem[]{}
Li, L.-X., Goodman, J., \& Narayan, R. 2003, ApJ, 593, 980.

\bibitem[]{}
Lin, D.N.C., Papaloizou, J. 1979, MNRAS, 186, 799

\bibitem[]{}
Lubow, S.H, 1981, ApJ, 245, 274

\bibitem[]{}
Lubow, S.H., Ogilvie, G.I., 1998, ApJ, 504, 983

\bibitem[]{}
Lubow, S.H., Pringle, J.E., 1993, ApJ, 409, 360

\bibitem[]{}
Masset, F.S., Papaloizou, J.C.B., 2003, ApJ, 588, 494

\bibitem[]{}
Meyer-Vernet, N., \& Sicardy, B. 1987, Icarus, 69, 157

\bibitem[]{}
Ogilvie, G.I., 2002, MNRAS, 331, 1053

\bibitem[]{}
Ogilvie, G.I., \& Lubow, S.H. 2003, ApJ, 587, 398

\bibitem[]{}
Okazaki, A., Kato, S., \& Fukue, J. 1987, PASJ, 39, 457

\bibitem[]{}
Olver, F.W.J. 1974, Asymptotics and Special Functions, Academic Press, New York

\bibitem[]{}
Papaloizou, J.C.B., \& Lin, D.C. 1995, ApJ, 438, 841


\bibitem[]{}
Shu, F.H., Yuan, C., Lissauer, J.J., 1985, ApJ, 291, 356

\bibitem[]{}
Takeuchi, T., \& Miyama, S.M. 1998, PASJ, 50, 141

\bibitem[]{}
Tanaka, H., Takeuchi, T., \& Ward, W.R. 2002, ApJ, 565, 1257 

\bibitem[]{}
Tanaka, H., \& Ward, W.R. 200, ApJ, 602, 388

\bibitem[]{}
Terquem, C. 1998, ApJ, 509, 819

\bibitem[]{}
Ward, W.R. 1986, Icarus, 67, 164

\bibitem[]{}
Ward, W.R. 1988, Icarus, 73, 330


\bibitem[]{}
Ward, W.R., 1997, ApJ, 482, L211

\bibitem[]{}
Ward, W.R., Hahn, J.M., 2003, ApJ, 125, 3389

\bibitem[]{}
Wong, R., 2001, Asymptotic Approximations of integrals, SIAM, Philadelphia


\end{thebibliography}
\end{document}